\documentclass[aps,
pre,
twocolumn,
superscriptaddress,
nofootinbib,
longbibliography
]{revtex4-2}
\usepackage[utf8]{inputenc}
\usepackage{hyperref}
\hypersetup{
	colorlinks=true,
	linkcolor=blue,
	filecolor=magenta,      
	urlcolor=blue,
	citecolor=blue,
	pdftitle={Overleaf Example},
	pdfpagemode=FullScreen,
}

\usepackage{graphicx}  
\usepackage{dcolumn}   
\usepackage{bm}        
\usepackage{amssymb,amsmath}   
\usepackage{uri}
\usepackage{float}
\usepackage{geometry}
\usepackage{comment}
\usepackage{fixmath}
\usepackage{makecell}
\begin{document}
	
	
	\title{Reentrant phase behavior in binary topological flocks with nonreciprocal alignment}

	\author{Tian Tang}
	\affiliation{National Laboratory of Solid State Microstructures and Department of Physics,\\
		Collaborative Innovation Center of Advanced Microstructures, Nanjing University, Nanjing 210093, China}
	\author{Yu Duan}
	\email{yu.duan@ds.mpg.de}
	\affiliation{Max Planck Institute for Dynamics and Self-Organization (MPI-DS), 37077 G\"ottingen, Germany}
	\author{Yu-qiang Ma}
	\email{myqiang@nju.edu.cn}
	\affiliation{National Laboratory of Solid State Microstructures and Department of Physics,\\
		Collaborative Innovation Center of Advanced Microstructures, Nanjing University, Nanjing 210093, China}
	\affiliation{Hefei National Laboratory, Hefei 230088, China}
	
	\date{\today}
	
	\begin{abstract}
		We study a binary metric-free Vicsek model involving two species of self-propelled particles aligning with their Voronoi neighbors, focusing on a weakly nonreciprocal regime, where species $A$ aligns with both $A$ and $B$, but species $B$ does not align with either.
		Using agent-based simulations, we find that even with a small fraction of $B$ particles, the phase behavior of the system can be changed qualitatively, which becomes reentrant as a function of noise strength: traveling bands arise not only near the flocking transition, but also in the low-noise regime, separated in the phase diagram by a homogeneous polar liquid regime.
		We find that the ordered bands in the low-noise regime travel through an \textit{ordered} background, in contrast to their metric counterparts.
		We develop a coarse-grained field theory, which can account for the reentrant phase behavior qualitatively, provided the higher-order angular modes are taken into consideration.
	\end{abstract}

	\maketitle
	\section{Introduction}
	
	Flocking, the emergent large-scale collective motion among locally aligning self-propelled particles~\cite{vicsek1995novel,chate2020dry,toner2024physics,solon2024thirty}, is a fascinating and ubiquitous self-organization phenomenon, arising in diverse living systems such as human crowds~\cite{bain2019dynamic}, bird flocks~\cite{ballerini2008interaction,cavagna2010scale}, bacterial colonies~\cite{nishiguchi2017long}, and driven filaments~\cite{schaller2010polar}, as well as in synthetic systems from active colloids~\cite{bricard2013emergence,das2024flocking}, to driven grains~\cite{deseigne2010collective,kumar2014flocking}, to robot swarms~\cite{vasarhelyi2018optimized}.
	Despite great complexity of real systems, flocking can be captured by the seminal Vicsek model (VM) in a minimal way~\cite{vicsek1995novel,chate2020dry}: particles are self-propelled along their headings, and align their headings with neighbors in competition with some noise.
	
	In addition to the exotic properties of the flocking itself~\cite{tu1998sound,geyer2018sounds,mahault2019quantitative,tasaki2020hohenberg,mahault2021long,fava2024strong}, the flocking transition, due to its historical and paradigmatic value, has remained a subject of intense studies~\cite{chate2020dry}.
	Flocking models can often be divided into two classes: the metric case, in which particles within a characteristic distance are recognized as neighbors~\cite{vicsek1995novel}, and the metric-free (or topological) case with more complex constructions of neighbor sets such as Voronoi or $k$-nearest neighbors ($k$NN)~\cite{ginelli2010relevance,martin2021fluctuation,Martin_2024,rahmani2021topological,shi2022collective,ito2024boltzmann}.
	The metric case is better understood, exhibiting a discontinuous transition between a disordered gas phase and a polar liquid (flocking) phase, separated in the phase diagram by a coexistence region where ordered bands travel through a disordered background~\cite{solon2015phase,chate2020dry}\footnote{
		Nevertheless, for metric flocks, we note there still leaves a door open for continuous transitions in certain situations, such as the case in the presence of quenched disorder~\cite{duan2021breakdown} or with delicate dependence of transport coefficients on density~\cite{bertrand2022diversity,jentsch2023critical,agranov2024thermodynamically}.}.
	
	On the other hand, topological flock models, which play an important role in research on groups of animals~\cite{ballerini2008interaction,Gautrais2012DecipheringII,camperi2012spatially,ginelli2015intermittent,pearce2014role,Jiang2017IdentifyingIN,jhawar2020noise} and pedestrians~\cite{moussaid2011simple}, are more challenging to study from both simulation and theory sides, due to the nonlocal properties of topological interactions.
	The flocking transition in this case was initially believed to be continuous, as first revealed by large-scale simulations of the Voronoi-VM~\cite{ginelli2010relevance} and subsequently rationalized by mean-field theories~\cite{peshkov2012continuous,chou2012kinetic}.
	However, recently this conclusion has been challenged~\cite{martin2021fluctuation,Martin_2024}.
	While the numerical evidence for discontinuous transitions in the $k$NN-VM is clear~\cite{martin2021fluctuation,Martin_2024}, the nature of transition in the Voronoi case still remains elusive~\footnote{Very recently, it was suggested that the transition is discontinuous in the continuous-time Voronoi-VM with small time-step sizes $\sim 10^{-2}$, but still appears continuous in simulations with millions of particles in the discrete-time case~\cite{packard2024banded}.}.
	
	Despite these on-going debates restricted to single-component systems, population heterogeneity, often accompanied with \textit{nonreciprocity}, is unavoidable in real flocking systems~\cite{maity2023spontaneous}. 
	In recent years, systems in which effective interactions violate the action-reaction symmetry have emerged as a paradigm for active matter~\cite{soto2014self,lavergne2019Science,fruchart2021non,saha2020scalar,you2020nonreciprocity,duan2023dynamical,dinelli2023non,chen2024emergent,kant2024bulk}.
	Many efforts therefore have been devoted to understanding the phase behavior of metric flocks with nonreciprocal interactions~\cite{chen2017fore,fruchart2021non,martin2023exact,maity2023spontaneous,kreienkamp2024non,mangeat2024emergent}, to name just a few.
	Somewhat surprisingly, little attention regarding nonreciprocity has been paid to topological flocks, even though the topological interactions are often vision-based, and are therefore typically nonreciprocal.
	Less is known about whether the nonreciprocal topological flocks can exhibit collective behavior qualitatively different from their metric counterparts.
	
	In this work, we study the two-species Voronoi-VM, focusing on a weakly nonreciprocal regime where only $A$ particles tend to align with their neighbors, while $B$ are dissenters and do not align with any~\cite{bera2020motile,yllanes2017many}.
	Using agent-base simulations, we find even a very small fraction of dissenters can change the system behavior qualitatively: the traveling bands not only emerge near the flocking transition, but also arise at low noise strength far below the transition, separated in the phase diagram by a polar liquid regime that disappears for large enough dissenter densities.
	Besides, we reveal that in the low-noise band regime, the ordered bands travel through an ordered background, in contrast to the metric case~\cite{solon2015phase}.
	Moreover, from the coarse-grained field theory, we obtain the linear stability diagram that can account for our simulation results.
	
	In the remainder of this paper, we first introduce the two-species nonreciprocal Voronoi-VM in Sec.~\ref{sec:micro-model}, where we also discuss the results of related metric models briefly.
    We then report the finite-size phase behavior of the system in Sec.~\ref{sec:finite-phase-behavior}, give an intuitive explanation for band formation in Sec.~\ref{sec:density-polarity}, and reveal the unusual properties of traveling bands at low noise, below the polar liquid, in Sec.~\ref{sec:abnormal-band}.
	To account for these simulation results, we derive a continuous theory in Sec.~\ref{sec:coarse-grained-continuous-theory} and discuss the results of linear stability analysis in Sec.~\ref{sec:linear-instability}.
	We finally provide concluding remarks and perspectives on our results in Sec.~\ref{sec:conclusion}.
	
	\section{Binary Voronoi-VM}\label{sec:micro-model}
	We consider a binary mixture of total $N$ particles moving at constant speed $v_0$.
	In 2D, particle $i$ of species $S\in\{A, B\}$ is defined by its position $\mathbf{r}_{i,S}^t$ and orientation $\theta_{i,S}^t$, updated at unit time steps according to 
	\begin{subequations} \label{eq:micro-model}
		\begin{align}
			\theta_{i,S}^{t+1}&=\arg \left(\sum_{j\sim i}J_{SS'}\hat{\mathbf{e}}(\theta_{j,S'}^{t})
			\right)+\eta \xi_i^t, \label{eq:micro-model-a}\\
			\mathbf{r}_{i, S}^{t+1}&=\mathbf{r}_{i, S}^{t} + v_0 \hat{\mathbf{e}}(\theta_{i,S}^{t+1}),      
		\end{align}
	\end{subequations}
	where summation is over all neighbors $j$ (from species $S'\in\{A, B\}$) of particle $i$,
	$J_{SS'}$ is the coupling strength between species $S$ and $S'$, $\hat{\mathbf{e}}(\theta)$ is the unit vector along $\theta$, and $\xi_i$ a uniform white noise drawing in $[-\pi, \pi]$, 
	while $\eta$ denotes the noise strength assumed equal for the two species.
	Unlike the metric case where neighbors $j$ are set by distance-based criteria, here they are chosen to be the $\mathcal{N}^t_i$ particles forming the first shell around particle $i$ in the Voronoi tessellation constructed from the particle positions at time $t$~\cite{ginelli2010relevance}.
    Excluded-volume interactions are ignored, as assumed in most numerical studies of flocking~\cite{chate2020dry}, which is especially justified in the metric-free case where the range of aligning interactions can be much larger than particle diameters.
	
	For the two-species case, nonreciprocity arises naturally when $J_{AB}\neq J_{BA}$.
	Despite strongly nonreciprocal interactions with $J_{AB}J_{BA} < 0$ postulated in the previous work for the metric case~\cite{fruchart2021non}, here we just consider weakly nonreciprocal couplings with $J_{AA}=J_{AB}=1$ but $J_{BA}=J_{BB}=0$, such that $A$ particles are aligners and $B$ dissenters.
    
	We note that similar aligner-dissenter mixtures have also been studied in the metric case both experimentally~\cite{bera2020motile} and by simulation~\cite{yllanes2017many}.
    It was reported that, upon increasing the fraction of dissenters, the order-disorder transition shifts to higher packing fractions of aligners~\cite{bera2020motile}, and dissenters can disrupt the flocking state more efficiently when excluded-volume interactions are dominated~\cite{yllanes2017many}.
    Indeed, the dissenters there are just expected to act like annealed disorder, without having a qualitative effect on the nature of the flocking transition and the polar liquid below this transition.
	However, in this work, we show that dissenters do play a nontrivial role in the metric-free models.
	
	In our simulations, we set self-propulsion speed $v_0=0.5$ and use square domains of linear size $L$ with periodic boundary conditions.
	The mean densities of $A$ and $B$ particles are denoted as $\bar{\rho}_A$ and $\bar{\rho}_B=\rho_0-\bar{\rho}_A$ respectively, with $\rho_0=N/L^2$ the total mean density.
	In the following, we set $\rho_0=1$ for simplicity. We are then left with two controlling parameters: the noise strength $\eta$ and the fraction of $B$ particles $\bar{\rho}_B/\rho_0=\bar{\rho}_B$. 
	
	\section{Phase behavior}\label{sec:simulation}
	\subsection{Finite-size phase behavior}\label{sec:finite-phase-behavior}
	\begin{figure*}
		\centering
		\includegraphics[width =\linewidth]{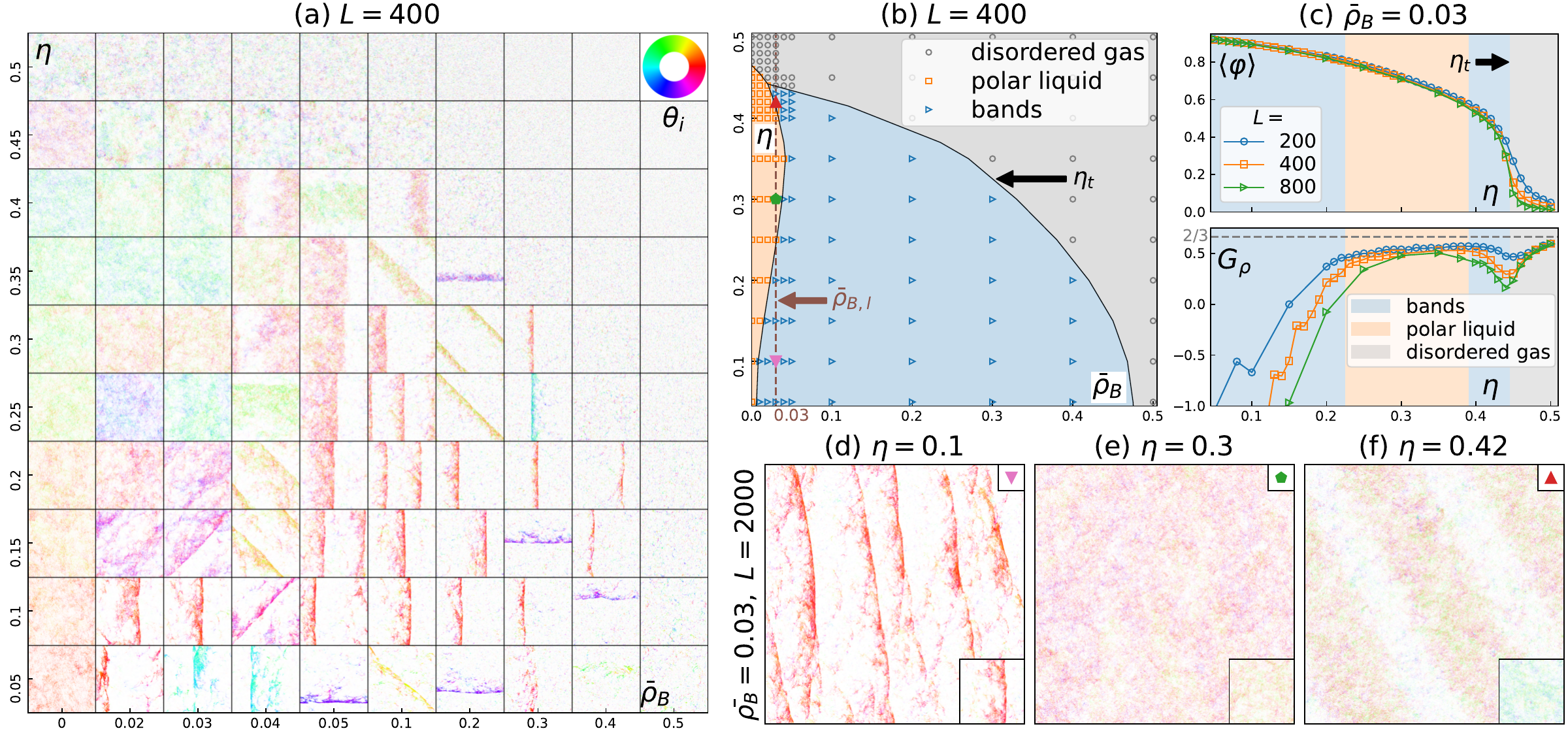}
		\caption{(a) Snapshots from simulations of the microscopic model~\eqref{eq:micro-model} at time $t=10^6$ in the $(\bar{\rho}_B, \eta)$ plane in square domains of size $L=400$. Particles from species $A$ are color-coded by their polarity orientation $\theta_i$ according to the inserted color disk, while those from species $B$ are not shown for clarity.
			(b) Finite-size phase diagrams in the $(\bar{\rho}_B, \eta)$. Hollow and solid symbols correspond to simulations with $L=400$ and $2000$ respectively. 
			(c) Order parameter $\langle \varphi\rangle$ and Binder cumulant of density fields $G_\rho$ as functions of $\eta$ with fixed $\bar{\rho}_B=0.03$ and varied $L$. The coarse-graining length scale to obtain the density fields is $\Delta x=40$.
			The phase boundaries are obtained by visual inspection of snapshots for $L=800$.
			(d-f) Detail of reentrant phase behavior at $\bar{\rho}_B=0.03$ as $\eta$ is increased, with $L=2000$ for the main panels and $L=400$ for the insets.
		}
		\label{fig:f1}
	\end{figure*}
	To understand the phase behavior of the binary topological flocks, we performed simulations of the microscopic model~\eqref{eq:micro-model} in periodic square boxes with size $L=400$, for varied $\bar{\rho}_B$ and noise strength $\eta$.
	Initialized with random particle positions and ordered headings, the simulation was run until $t=10^6$ to ensure good statistics for the measurement discussed below.
	
	The phase behavior of the system is illustrated in Fig.~\ref{fig:f1}(a) by typical snapshots and in Fig.~\ref{fig:f1}(b) by the finite-size phase diagram in the $(\bar{\rho}_B, \eta)$ plane, where the phase boundaries are delineated by visual inspection of snapshots.
	
	As shown in Fig.~\ref{fig:f1}(a, b), at $\bar{\rho}_B=0$, there is a direct transition from polar liquid to disordered gas with increasing $\eta$, in line with the previous study for the one-species case~\cite{ginelli2010relevance}.
	
	As $\bar{\rho}_B$ is increased from $0$, the order-disorder (or flocking) transition threshold $\eta_t$ is decreased as expected.
	Interestingly, near the order-disorder transition ($\eta\sim \eta_t$), bands start to emerge for $\bar{\rho}_B\geq \bar{\rho}_{B, l}\approx0.03$ (brown dashed line in Fig.~\ref{fig:f1}(b)).
	On the other hand, surprisingly, even before $\bar{\rho}_B$ is increased to $\bar{\rho}_{B, l}$, bands already arise in the low-noise regime with $\eta$ far below $\eta_t$.
	Hence, the phase behavior becomes reentrant at $\bar{\rho}_B=0.03$, where the two band regimes are separated by a polar liquid regime.
	The latter shrinks with increasing $\bar{\rho}_B$, and eventually disappears when $\bar{\rho}_B$ goes beyond $\bar{\rho}_{B,h}\approx 0.05$, leaving only the band phase and the disordered gas separated by $\eta_t(\bar{\rho}_B)$, which is decreased to zero at $\bar{\rho}_B\approx 0.48$. 
	Afterwards, the system is always disordered for any $\eta$.
	
	To quantify these transitions, we calculate the time-average order parameter $\langle \varphi \rangle$ for $A$ particles and the Binder cumulant $G_\rho\equiv 1 - \langle \rho_A(\mathbf{r})^4\rangle/3\langle\rho_A(\mathbf{r})^2\rangle^2$ of density $\rho_A(\mathbf{r})$, where $\varphi(t)\equiv |\langle \exp(i\theta_{j, A}^t) \rangle_{j}|$.
	For $\rho_A(\mathbf{r})$ obtained by coarse-graining with suitable length scales, one expects $G_\rho\sim 2/3$ in the homogeneous phase and deviates from $2/3$ notably in the band phase.
	As shown in Fig.~\ref{fig:f1}(c), for fixed $\bar{\rho}_B=0.03$, with increasing $\eta$, there is a rapid decrease in $\langle\varphi\rangle$ as $\eta\to\eta_t$, signaling an order-disorder transition there, which becomes sharper and sharper with increasing $L$.
	Meanwhile, $G_\rho$ varied non-monotonically with $\eta$, confirming the reentrant phase behavior previously identified by visual inspection of snapshots (Fig.~\ref{fig:f1}(a, b))

	We further performed several simulations with $L=2000$ to probe the large-scale phase behavior.
	The resulting snapshots are shown in Fig.~\ref{fig:f1}(d-f), confirming the existence of reentrant phase behavior in such a large system (see also Supplemental Movie (SMov) 1 for the $L=1000$ case).
	Note that while the polar liquid at $\eta=0.3$ remains nearly unchanged with increasing $L$ (Fig.~\ref{fig:f1}(e) and its inset), near the order-disorder transition, the polar liquid seen in small systems is superseded by the band phase when $L$ gets large enough (Fig.~\ref{fig:f1}(f) and its inset).
	This suggests that the lower density threshold $\bar{\rho}_{B, l}\approx 0.03$, beyond which the band can arise near the order-disorder transition, would decrease with increasing $L$.
	In the thermodynamic limit, it is likely that $\bar{\rho}_{B,l}$ could approach zero such that the order-disorder transition is discontinuous for arbitrary small $\bar{\rho}_{B}>0$.
	
	\subsection{Density-polarity coupling}\label{sec:density-polarity}
	
	Having established the existence of band phase in the two-species case (Fig.~\ref{fig:f1}), we now give an intuitive explanation of why a small fraction of $B$ particles can affect the system behavior so dramatically.
	A more comprehensive theoretical consideration, which can qualitatively reproduce the phase diagram shown in Fig.~\ref{fig:f1}(b), will be presented in Sec.~\ref{sec:cont-description}.
	
	The arising of band phase in VM can ultimately be attributed to the positive feedback mechanism between the local density and polarity~\cite{bertin2009hydrodynamic,mishra2010fluctuations}: if an increase in the local density enhances the local polarity sufficiently, then particles belonging to the denser cluster would move more coherently, recruiting more particles from the surrounding background of lower polarity (and also of low density), which in turn enhances local densities of the denser cluster, so that the denser cluster would continue growing, eventually destabilizing the homogeneous phase.
	
	For metric-free interactions, increasing the local density does not increase the number of neighbors with which a particle aligns.
	When all particles are identical aligners, the local polarity and density are naturally decoupled in the mean-field limit~\cite{peshkov2012continuous,chou2012kinetic}.
	In contrast, for fixed $\bar{\rho}_B>0$, the higher $\rho_A(\mathbf{r})$ is, the less likely $A$ particles are to take $B$ as their neighbors, so that the local polarity will increase with increasing local $A$ density.
	
\begin{figure}
		\centering
		\includegraphics[width =\linewidth]{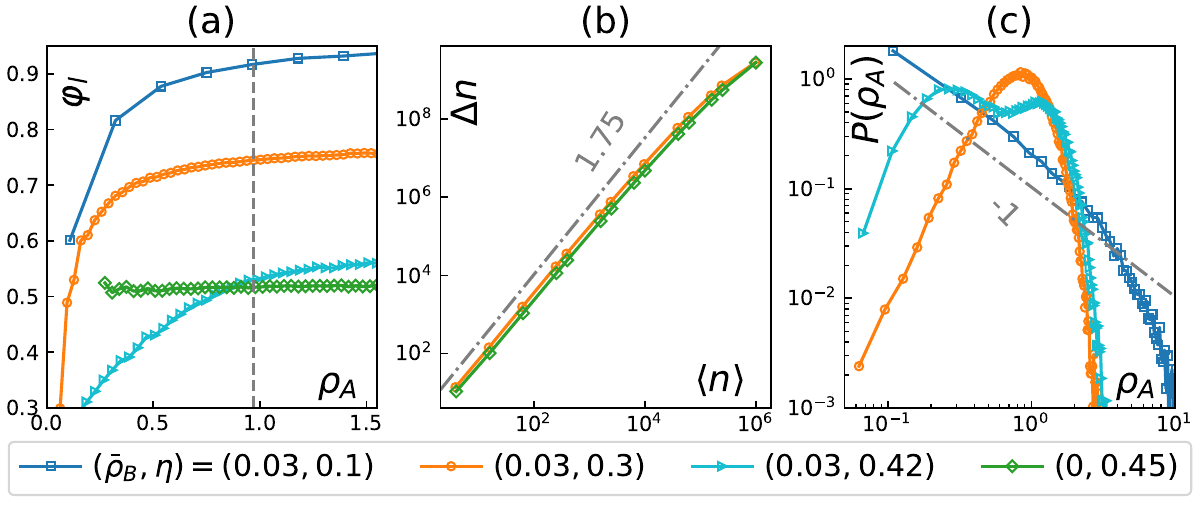}
		\caption{
			Statistics of coarse-grained fields at different $(\bar{\rho}_B,\eta)$ with $L=2000$.
			Blue, orange and cyan symbols correspond to the phases shown in Fig.~\ref{fig:f1}(d-f) respectively, while green ones correspond to the polar liquid near the order-disorder transition with $\bar{\rho}_B=0$.
			(a) Local order parameter $\varphi_l$ versus local density $\rho_A$ with coarse-graining length scale $\Delta x=40$.
			(b) Giant number fluctuations in the polar liquid phase: variance of particle number $\Delta n$ versus mean particle number $\langle n\rangle$ where $n$ is the particle number of species $A$ contained in boxes of various linear sizes.
                (d) Probability distribution function of coarse-grained density $\rho_A(\mathbf{r})$ with $\Delta x=40$.
		}
		\label{fig:f2}
	\end{figure}

	To quantify the polarity-density coupling, with the local density $\rho_A(\mathbf{r})$ and local order parameter $\varphi(\mathbf{r})$, we can further bin appropriately the range of $\rho_A$, and for each bin compute the average value of the corresponding $\varphi(\mathbf{r})$, obtaining the local order parameter $\varphi_l$ as a function of the local density $\rho_A$.
	For $\bar{\rho}_B=0$, in the polar liquid phase close to the order-disorder transition, we find a very weak dependence of $\varphi_l$ on $\rho_A$ as expected (green curve in Fig.~\ref{fig:f2}(a)).
	In contrast, $\varphi_l$ increases with $\rho_A$ notably in the other three cases with $\bar{\rho}_B=0.03$ shown in Fig.~\ref{fig:f2}(a).
	More precisely, in the band phases (blue and cyan curves in Fig.~\ref{fig:f2}(a)), the local slopes of the $\varphi_l(\rho_A)$ curves at $\rho_A=\bar{\rho}_A=0.97$ (gray dashed line in Fig.~\ref{fig:f2}(a)) are larger than those in the polar liquid phase (orange curve in Fig.~\ref{fig:f2}(a)), in consistent with the intuition that the stronger the polarity-density coupling is,  the more likely the polar liquid could be destabilized.
	
	\subsection{Abnormal bands}\label{sec:abnormal-band}
	\begin{figure}
		\centering
		\includegraphics[width =\linewidth]{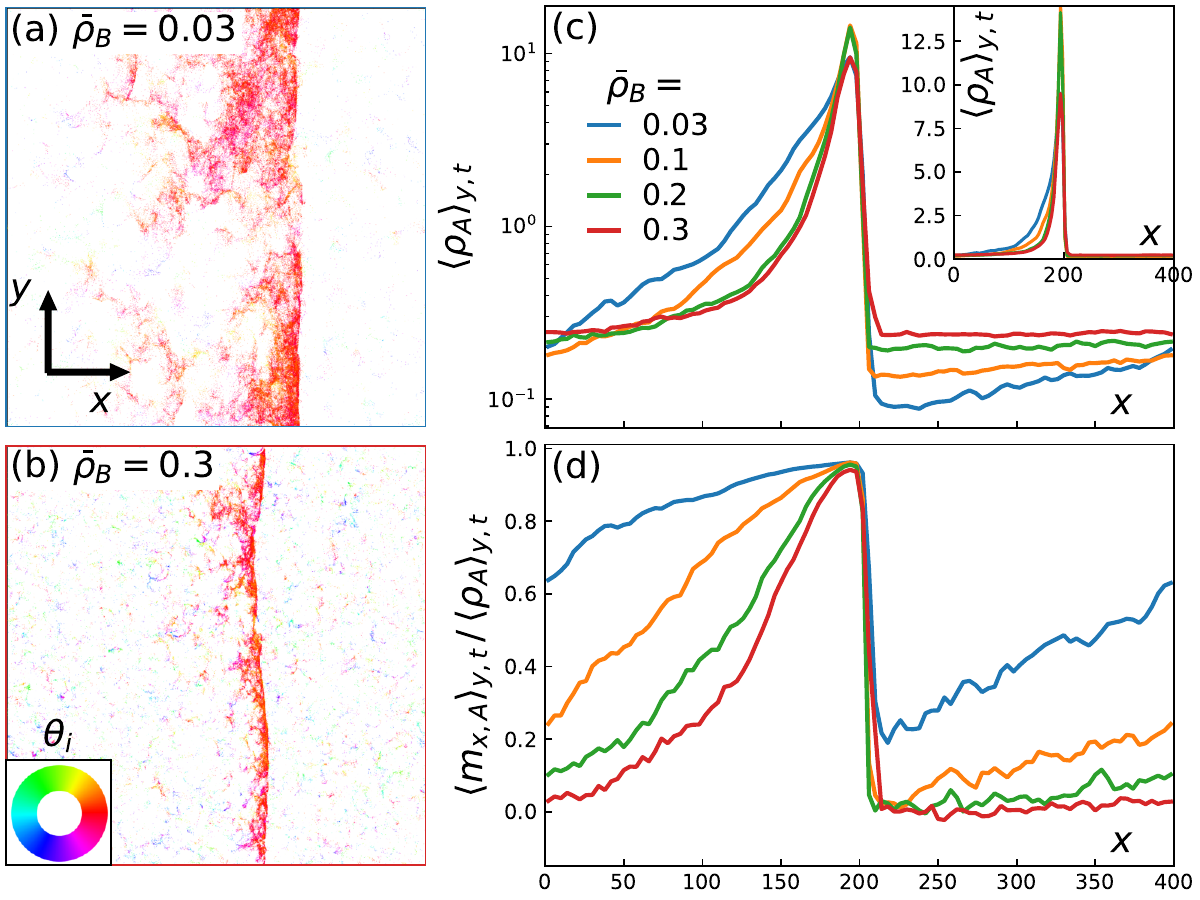}
		\caption{Crossover from the abnormal to normal band regime with fixed $\eta=0.1$ and increasing $\bar{\rho}_B$.
			(a, b) Typical snapshots for a single band moving along the $x$ direction in square domains of size $L=400$, with $\bar{\rho}_B=0.03$ and $0.3$ respectively.
			(c) $y-$ and time-averaged density $\langle \rho_A(\mathbf{r},t) \rangle_{y, t}$ versus $x$ in lin-log scales with varied $\bar{\rho}_B$. Inset: the same plot as (c) but in lin-lin scales.
			(d) Polarity profiles $\langle m_{x,A}\rangle_{y, t} / \langle \rho_A\rangle_{y, t}$ for varied $\bar{\rho}_B$, where $m_{x,A}(\mathbf{r},t)\equiv \sum_i \delta[\mathbf{r}-\mathbf{r}_{i, A}(t)] \cos [ \theta_{i, A}(t)]$ is the $x$ component of momentum fields for $A$ particles.
		}
		\label{fig:f3}
	\end{figure}
    Although for $\bar{\rho}_B>0$, the polar liquid becomes unstable in a large portion of the phase diagram (Fig.~\ref{fig:f1}(b)), the survived polar liquid still exhibits giant number fluctuations (GNF) with a scaling exponent $\approx 1.75$ (orange curve in Fig.~\ref{fig:f2}(b)), the same as the one-species case (green curve in Fig.~\ref{fig:f2}(b))~\cite{ginelli2010relevance}, suggesting the presence of $B$ particles would not cause a qualitative change to the collective properties of the polar liquid that survives.
    
    However, the two band regimes separated by the polar liquid in the phase diagram (Fig.~\ref{fig:f1}(b)) possess distinct properties from each other.
	At first glance, the two types of bands shown in Fig.~\ref{fig:f1}(d) and (f) appear similar, albeit with different band widths that may be selected by fluctuations~\cite{solon2015phase}.
	However, a qualitative difference between the two cases appears when we look at the probability distribution functions (PDFs) of $\rho_{A}(\mathbf{r})$, which are unimodal in the polar liquid phase, bimodal in the high-noise band regime, but decay monotonically with $\rho_A$ in the low-noise band regime (Fig.~\ref{fig:f2}(c)), suggesting that density fields $\rho_{A}(\mathbf{r})$ are highly inhomogeneous in the last case.
	
	Looking closely at the one-band configurations, we find that in the low-noise band regime, the band has a long tail and the surrounding gas seems to be ordered (Fig.~\ref{fig:f3}(a)).
	This abnormal behavior is confirmed by the corresponding density and polarity profiles, revealing that the gas phase surrounding the band is indeed ordered and inhomogeneous (blue curves in Fig.~\ref{fig:f3}(c, d)). 
        On the other hand, we find normal traveling bands near the order-disorder transition (Fig.~\ref{fig:f3}(b)), for which the coexisting gas phase is homogeneous and disordered (red curves in Fig.~\ref{fig:f3}(c, d)).
	With increasing $\bar{\rho}_B$, the local polarity in the coexisting gas phase decreases, and the abnormal band crosses over to the normal one (Fig.~\ref{fig:f3}(c, d)).
	
	Given that there are more bands with larger $L$ (Fig.~\ref{fig:f1}(d) and its inset), the abnormal bands at the low-noise regime appear to be a state involving micro-phase separation between polar liquid and polar gas.
	This certainly violates the established picture of phase coexistence in polar flocks~\cite{chate2020dry}, where the bands travel through a disordered gas~\cite{solon2015phase}.
	We note that phase separation between two ordered phases of different densities has been predicted by previous works~\cite{bertrand2022diversity,miller2024following} based on phenomenological Toner-Tu equations~\cite{toner2024physics}, but the phase separation there is macroscopic and should result from very different physical mechanisms to us.
    Given that the coexisting gas phase remains inhomogeneous, the abnormal bands here could be more relevant to the limit cycle of the hydrodynamic equation~\cite{solon2015pattern} or the Boltzmann equation~\cite{mahault2018outstanding} derived from the one-species metric VM, although in simulations of this microscopic model, similar bands traveling through an ordered gas have not been observed instead~\cite{chate2020dry}.

	\section{Continuous description}\label{sec:cont-description}
	To better understand the physical mechanisms underlying the rich phase behavior observed in simulations (Fig.~\ref{fig:f1}), we derive a mean-field theory by coarse-graining the microscopic model (Sec.~\ref{sec:coarse-grained-continuous-theory}).
	We find that to account for the phase diagram shown in Fig.~\ref{fig:f1}(b), instead of the hydrodynamic equations (Sec.~\ref{sec:lin-hydro}), we need to resort to the kinetic equations (Sec.~\ref{sec:lin-kinetic}).
	
	\subsection{Coarse-grained continuous theory}\label{sec:coarse-grained-continuous-theory}
	As the aligning interactions involved in Eq.~\eqref{eq:micro-model-a} are of multi-body nature, it is difficult to coarse-grain the microscopic model~\eqref{eq:micro-model} directly to obtain the corresponding continuous description\footnote{
		Nevertheless, we notice it is still possible, albeit somewhat cumbersome, to do so
		by means of an Enskog-type kinetic theory which deals with multi-body interactions explicitly~\cite{chou2012kinetic}.
	}.
	Here, we follow a lightweight Boltzmann approach with the assumption
	that aligning ``collisions” occur at a low rate such that they are dominated by binary interactions~\cite{peshkov2012continuous,peshkov2014boltzmann}.
	In this way, well-controlled nonlinear field equations have been obtained for the one-species metric-free case~\cite{peshkov2012continuous}.
	For completeness, in Appendix~\ref{app:model-low-rates} we directly study a microscopic model with low ``collision" rates, in which we find similar reentrant phase behavior.
	This indicates that the two models, with and without low collision rates respectively, share the same collective properties, and thus the continuous theory derived from the former should also be applicable to the latter.

	Following~\cite{peshkov2012continuous}, by a straightforward generalization to the two-species case as detailed in Appendix~\ref{app:deriv-eq}, we get the Boltzmann equations about the one-particle phase-space distribution $f_S(\mathbf{r},\theta, t)$ for species $S\in\{A, B\}$, whose angular Fourier modes
	$f_{k,S}(\mathbf{r}, t)=\int_{-\pi}^\pi \mathrm{d}\theta f_S(\mathbf{r},\theta, t)e^{ik\theta}$
	evolve according to
	\begin{subequations}\label{eq:kinetic}
		\begin{align}
			\partial_t f_{k, A} &= -\frac{1}{2} (\triangledown f_{k-1, A}+\triangledown^* f_{k+1, A}) + (P_k-1-\alpha)f_{k, A} \notag \\
			&+ \frac{\alpha}{\rho} P_k\sum_{l=-\infty}^\infty I_{kl} f_{k-l, A} (f_{l, A}+f_{l, B}),\label{eq:kinetic-a} \\ 
			\partial_t f_{k, B} &= -\frac{1}{2} (\triangledown f_{k-1, B}+\triangledown^* f_{k+1, B}) + (P_k-1)f_{k, B} \label{eq:kinetic-b}
		\end{align}
	\end{subequations}
	where we have used complex operators $\triangledown =\partial_x + i\partial_y$ and $\triangledown^* =\partial_x -i\partial_y$, while $\rho(\mathbf{r},t)=\rho_A(\mathbf{r},t)+\rho_B(\mathbf{r},t)$ denotes total density fields of the two species,
	$P_k(\eta)=\exp(-k^2 \eta^2/2)$ depends on noise strength $\eta$,
	and
	$I_{kl}$ is an integral whose explicit expression is given by Eq.~\eqref{eq:I_kl} in Appendix~\ref{app:deriv-eq}.
	Here, $\alpha$ denotes the collision rate that has been expressed in the rescaled units, and is therefore not necessarily smaller than one~\cite{peshkov2012continuous}. In the following, we set $\alpha=1$ for simplicity, while we have checked that our results still hold for other $\alpha$ values.
	
	Noting that $f_{0, S}=\rho_S$, Eq.~\eqref{eq:kinetic} is therefore left unchanged when normalizing $f_{k, A}$ and $f_{k, B}$ by an arbitrary factor for all $k$ simultaneously. This indicates that the dynamics of $f_{k,S}$ is independent of the total mean density $\rho_0$, a manifestation of the basic properties of metric-free interactions encoded in the microscopic model~\eqref{eq:micro-model}.
	
	Obviously, the solution for Eq.~\eqref{eq:kinetic-b} is
	\begin{equation}\label{eq:solution-f_k_B}
		\rho_B(\mathbf{r}, t) = \bar{\rho}_B,\quad f_{k, B}(\mathbf{r}, t) = 0\quad \mathrm{for}\quad k > 0,
	\end{equation}
	since $B$ particles just move randomly, yielding uniform density fields and vanishing angular modes for $k\geq 1 $.
	
	Substituting solution~\eqref{eq:solution-f_k_B} into Eq.~\eqref{eq:kinetic-a}, and utilizing the scaling ansatz
	\begin{equation}\label{eq:scaling-ansatz}
		\rho_A-\bar{\rho}_A\sim\varepsilon,\quad f_{k, A}\sim \varepsilon^{|k|},\quad \triangledown\sim \varepsilon,\quad \partial_t\sim \varepsilon,
	\end{equation}
	which is valid near onset of polar order~\cite{peshkov2014boltzmann}, we truncate Eq.~\eqref{eq:kinetic-a} at the first non-trivial order $\varepsilon^3$, obtaining the equations for $\rho_A$, $f_{1, A}$ and $f_{2, A}$. 
	Taking the quasi-stationary approximation for $f_{2, A}$ and enslaving it to the lower order modes, we find the closed equations for $\rho_A$ and $f_{1, A}$, which read after going back to vector notations
	\begin{subequations} \label{eq:hydro}
		\begin{align}
			\partial_t \rho_A &+ \nabla \cdot \mathbf{w}_A = 0, \label{eq:hydro-a} \\
			\partial_t \mathbf{w}_A &+ \gamma(\mathbf{w}_A\cdot \nabla)\mathbf{w}_A = -\frac{1}{2} \nabla \rho_A+\frac{\kappa}{2}\nabla \mathbf{w}_A^2 \label{eq:hydro-b}\\
			&+ (\mu-\xi \mathbf{w}_A^2)\mathbf{w}_A + \nu\nabla^2\mathbf{w}_A - \kappa(\nabla\cdot \mathbf{w}_A)\mathbf{w}_A, \notag
		\end{align}
	\end{subequations}
	where $\mathbf{w}_A\equiv(\Re(f_{1,A}), \Im(f_{1,A})))$ denotes the momentum vector of species $A$.
	Eq.~\eqref{eq:hydro} shares the same form as those derived in the one-species case for both metric-free and metric interactions~\cite{peshkov2012continuous,bertin2006boltzmann}, but with different transport coefficients.
	
	Here we just need to notice that in Eq.~\eqref{eq:hydro-b}, while the other coefficients $\gamma$, $\nu$, $\kappa$ and $\xi$ are always positive (see Eq.~\eqref{eq:coefficients-hydro} in Appendix~\ref{app:deriv-eq} for their expressions), the coefficient of the linear term, $\mu$, can change sign as seen from its expression
	\begin{equation}\label{eq:mu}
		\mu = \left(
		\frac{2\alpha(2{\rho}_A+\bar{\rho}_B)}{\pi ({\rho}_A+\bar{\rho}_B)} + 1
		\right) P_1(\eta) - (1+\alpha),
	\end{equation}
	where $P_1(\eta)=\exp(-\eta^2/2)$ decreases with increasing noise strength $\eta$.
	In Fig.~\eqref{eq:hydro-b}, given that $\xi > 0$, the homogeneous disordered solution $(\rho_A, \mathbf{w}_A)=(\bar{\rho}_A, 0)$ is stable if $\mu$ is negative.
	By setting $\rho_A$ to $\bar{\rho}_A=\rho_0-\bar{\rho}_B$, from Eq.~\eqref{eq:mu} we find $\mu<0$ requires $\eta>\eta_t$, where $\eta_t$, the noise strength threshold for the order-disorder transition, is defined by
	\begin{equation} \label{eq:eta_t}
		\eta_t = \begin{cases}
			\sqrt{2\ln \left(
				\frac{\pi + 4\alpha - 2\alpha \bar{\rho}_B/\rho_0 }{\pi(1+\alpha)}
				\right)}, &\mathrm{if}\ \frac{\bar{\rho}_B}{\rho_0} < \frac{4-\pi}{2}, \\
			0, &\mathrm{if}\ \frac{\bar{\rho}_B}{\rho_0} \geq \frac{4-\pi}{2}.
		\end{cases}
	\end{equation}
	If $\bar{\rho}_B/\rho_0> (4-\pi)/2$, then $\eta_t=0$ and the system is always in the disordered phase. 
	Otherwise, $\eta_t$ becomes positive and for $\eta < \eta_t$ the homogeneous ordered solution $(\rho_A, \mathbf{w}_A)=(\bar{\rho}_A, \bar{\mathbf{w}}_A)$ exists where $\bar{\mathbf{w}}_A$ is a vector with magnitude $\sqrt{\mu/\xi}$ and arbitrary direction.
	In the $(\bar{\rho}_B,\eta)$ plane with fixed $\rho_0$, Eq.~\eqref{eq:eta_t} thus defines an order-disorder transition line $\eta_t(\bar{\rho}_B)$ that decreases with increasing $\bar{\rho}_B$ and eventually intersects the $\bar{\rho}_B$ axis, in line with the simulation results shown in Fig.~\ref{fig:f1}(a, b).
	
	\subsection{Linear stability analysis}\label{sec:linear-instability}
	The homogeneous ordered solution $(\bar{\rho}_A,\bar{\mathbf{w}}_A)$ that exists for $\eta < \eta_t$ is stable to homogeneous perturbations, but could be destabilized by the linear instability with respect to finite-length wave vectors $\mathbf{q}$.
	To examine the linear stability, we denote by $\delta\rho$ the fluctuations in $\rho_A$, while using $\delta w_A$ and $\delta n_A$ to represent the fluctuations in magnitudes and orientations of $\mathbf{w}_A$ respectively.
	Introducing $\mathbf{u}(\mathbf{r}, t)=(\delta \rho_A, \delta w_A, \delta n_A)$, we find the evolution of fluctuations $\mathbf{u}$ in Fourier space is dictated by
	\begin{equation} \label{eq:lin-hydro-M}
		\partial_t\mathbf{u}_\mathbf{q} (t) = \underline{M}(\mathbf{q})\mathbf{u}_\mathbf{q}(t),
	\end{equation}
	where $u_\mathbf{q}$ is the Fourier modes of $\mathbf{u}$ and $\underline{M}(\mathbf{q})$ is a $3\times 3$ matrix whose explicit expression is given by Eq.~\eqref{eq:M} in Appendix~\ref{app:linear-instability-hydro}, where we also show the details leading to Eq.~\eqref{eq:lin-hydro-M}.
	
	The stability of the homogeneous ordered solution $(\bar{\rho}_A,\bar{\mathbf{w}}_A)$ is therefore controlled by the eigenvalues of the matrix $\underline{M}$.
	Although the solution of the full cubic eigenvalue problem can be found numerically and will be discussed later, it is instructive to consider the case near the order-disorder transition first, at which the eigenvalue problem can be reduced to a quadratic one, allowing us to gain some analytical insights.
	
	\subsubsection{Linear instability near the order-disorder transition}
	Near the order-disorder transition, we show in Appendix~\ref{app:near-order-disorder} that dynamics of $\delta n_A$ is decoupled from those of $\delta \rho_A$ and $\delta w_A$, so that we only need to consider the coupled equations about the latter two.
	Solving the corresponding quadratic eigenvalue problem and denoting by $\sigma$ the growth rates of perturbations, we find the imaginary parts of $\sigma$ are always nonzero, while their largest real part becomes positive in the $q\to0$ limit iff (see Appendix~\ref{app:near-order-disorder} for the derivation)
	\begin{equation}\label{eq:instab-condi-1D}
		[\mu'-\xi' \mu/\xi - \gamma \mu]^2 > \gamma^2 \mu^2 + 2\mu \xi,
	\end{equation}
	where $\mu, \gamma, \xi$ are evaluated at mean density $\bar{\rho}_A$ and primes indicate derivation with respect to $\rho_A$.
	When condition~\eqref{eq:instab-condi-1D} is true, the homogeneous ordered solution becomes unstable with respect to long-wave oscillatory instability.
	
	Immediately, we find near the disorder-order transition, i.e., as $\eta\to\eta_t$ and $\mu\to 0$, condition~\eqref{eq:instab-condi-1D} is fulfilled once $\mu'\neq 0$.
	For the one-species case, i.e., with $\bar{\rho}_B=0$, Eq.~\eqref{eq:mu} leads to $\mu'=0$, therefore no instability can take place there at the mean-field level.
	However, $\mu'$ becomes nonzero once $\bar{\rho}_B>0$. 
	Hence, as $\eta\to\eta_t$, instability will arise in the two-species case.
	As $\eta$ is decreased from $\eta_t$, Eq.~\eqref{eq:instab-condi-1D} remains true until $\eta=\eta_{\parallel}$, the lower threshold for the instability, below which homogeneous ordered solutions become stable.
	
\begin{figure}
		\centering
		\includegraphics[width =\linewidth]{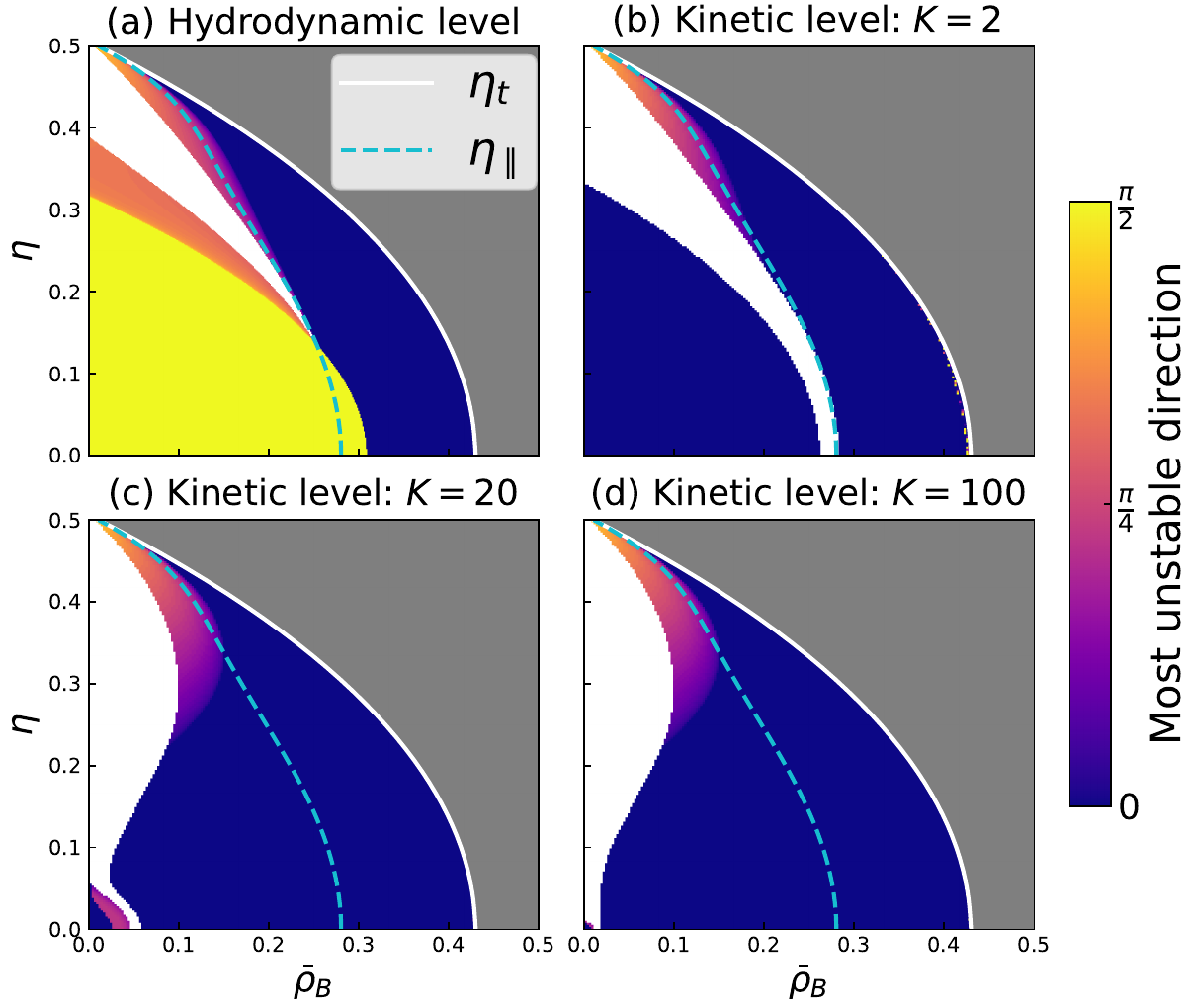}
		\caption{Linear stability diagrams obtained from (a) hydrodynamic equations~\eqref{eq:hydro} with linearized form~\eqref{eq:lin-hydro-M} and (b-d) kinetic equations~\eqref{eq:kinetic} with varied truncation orders $K$.
			Grey and white regions correspond to stable homogeneous disordered and ordered phase respectively, while linear instability regions are color-coded by the direction of the most unstable modes w.r.t.~the direction of the global polar order.
			The $\eta_t$ and $\eta_\parallel$ lines predicted by~\eqref{eq:instab-condi-1D} mark the region of instability arising from the linear system~\eqref{eq:lin-hydro-rho-w} where we only consider coupled dynamics between perturbations in density and momentum magnitude of $A$ particles.
		}
		\label{fig:f4}
	\end{figure}

	In Fig.~\ref{fig:f4}, we show the curves of $\eta_t(\bar{\rho}_B)$ and $\eta_\parallel(\bar{\rho}_B)$ in the $(\bar{\rho}_B,\eta)$ plane, between which the instability occurs.
	This explains the simulation finding that bands can arise near the order-disorder transition, but cannot account for the low-noise band regime shown in Fig.~\ref{fig:f1}.
	This discrepancy indicates that the dynamics of $\delta n_A$ becomes important as $\eta$ deviates from $\eta_t$, which we will consider in the next section.
	\subsubsection{Spurious linear instability for hydrodynamic equations}\label{sec:lin-hydro}
	Now we solve numerically the cubic eigenvalue problem for the linear system~\eqref{eq:lin-hydro-M}, taking into account the coupled dynamics of $\delta\rho$, $\delta w_A$ and $\delta n_A$.
    Following the procedure detailed in Appendix~\ref{app:num-3rd}, we get the linear stability diagram shown in Fig.~\ref{fig:f4}(a).
	Remarkably, now a low-noise instability regime appears, in which the most unstable modes are perpendicular to the direction of the global polarity (yellow region in Fig.~\ref{fig:f4}(a)).
	However, this instability still exists for a wide range of $\eta$ at $\bar{\rho}_B=0$, at which no bands can arise in simulations (Fig.~\ref{fig:f1}).
	
	Note that even in the absence of $B$ particles, in both metric and metric-free cases, one also finds a similar spurious instability at low noise far below the order-disorder transition~\cite{peshkov2014boltzmann,peshkov2012continuous}.
    This is inconsistent with the simulation results of microscopic models, for which the polar liquid is always stable in the corresponding parameter regime~\cite{peshkov2014boltzmann,peshkov2012continuous}.
	The existence of such spurious instability should be an artifact of the truncation ansatz~\eqref{eq:scaling-ansatz} which is no longer valid far below the order-disorder transition.
	Although it remains elusive how to get the well-behaved hydrodynamic equations in this situation, one can nevertheless perform linear instability analysis on the kinetic equations~\eqref{eq:kinetic} directly~\cite{mahault2018outstanding}, as we show below.
	
	\subsubsection{Linear instability for kinetic equations}\label{sec:lin-kinetic}
	To get round the spurious instability that arises at hydrodynamic level (Fig.~\ref{fig:f4}(a)), we turn to performing the linear stability analysis for the kinetic equations~\eqref{eq:kinetic} directly. 
	After truncating the infinite hierarchy~\eqref{eq:kinetic-a} at a given order $K$ by setting $f_{k, A}=0$ for all $|k|>K$, we get $K+1$ closed kinetic equations for $\{f_{k, A}\}_{0\leq k\leq K}$.
	Then we can study the stability of the homogeneous solution for these equations, following the procedure detailed in Appendix~\ref{app:linear-instability-kinetic}.
	The resulting linear stability diagrams for $K=2$, $20$ and $100$ are shown in Fig.~\ref{fig:f4}(b-d) respectively.
	
	As expected, for $\eta > \eta_t$ the homogeneous disordered solution is always stable (gray regions in Fig.~\ref{fig:f2}).
	Considering $\eta< \eta_t$, for truncation orders $K=2$, there are two instability regimes separated by a homogeneous ordered regime (Fig.~\ref{fig:f4}(b)).
	With increasing $K$, the instability regime on the left-hand side of the homogeneous ordered regime shrinks, and almost vanishes at $K=100$ (Fig.~\ref{fig:f4}(d)).
	On the other hand, the other instability regime, bordered by the $\eta_t(\bar{\rho}_B)$ curve, tends to expand as $K$ is increased, but becomes saturated for large enough $K$ (Fig.~\ref{fig:f4}(b-d)). Remarkably, at $K=100$ the linear stability diagram shown in Fig.~\ref{fig:f4}(d) quite resembles the phase diagram obtained in simulations (Fig.~\ref{fig:f1}(b)). 
	Furthermore, Fig.~\ref{fig:f4}(d) shows that near the order-disorder transition line $\eta_t(\bar{\rho}_B)$, the instability region shrinks with decreasing $\bar{\rho}_B$, and vanishes at $\bar{\rho}_B=0$.
	This is in line with our previous conjecture that the order-disorder transition would be discontinuous once $\bar{\rho}_B>0$ in the thermodynamic limit (Sec.~\ref{sec:finite-phase-behavior}).
	
	We therefore conclude that the kinetic equations~\eqref{eq:kinetic} can qualitatively account for the phase behavior of the microscopic model~\eqref{eq:micro-model}, while the hydrodynamic equations~\eqref{eq:hydro}, whose derivation relies on the scaling ansatz~\eqref{eq:scaling-ansatz}, do not work well in the low-noise regime.
	We note that higher-order orientation modes also play an important role in modeling nonreciprocal quorum-sensing active particles~\cite{duan2023dynamical,duan2024phase}.
	In this model, the continuous equations involving both density and polarity fields reveal a new instability regime, which is in line with particle simulation but is not predicted by the equations involving only density fields~\cite{duan2023dynamical,duan2024phase}.
	Nevertheless, it is difficult to gain analytical insight into the kinetic equations involving many higher-order modes.
	How to properly enslave the higher-order modes to the lower-order ones in the parameter regime far away from the onset of long-wave instability remains an open challenge~\cite{peshkov2014boltzmann} on which progress has been done recently~\cite{seyed2016gaussian}.
	
	\section{Conclusion and outlook}\label{sec:conclusion}
	
	In this work, we have investigated the phase behavior of two-species self-propelled particles aligning with their Voronoi neighbors in a weakly nonreciprocal regime, where $A$ particles are aligners and $B$ dissenters.
	Despite the simplicity of the microscopic dynamics, we have revealed, both via numerical simulations and the study of continuous models derived from coarse-graining, that even a very small fraction of dissenters can change the phase behavior of the system qualitatively: traveling bands, which are absent in the one-species case, start to arise, not only near the flocking transition, but also in the low-noise regime, with a polar liquid regime in between.
	The bands in the low-noise regime, below the polar liquid, are abnormal as they travel through an ordered background. With increasing the mean density of the dissenters, the abnormal bands cross over to the normal ones, accompanied by the shrinking of the polar liquid regime in the phase diagram, which eventually vanishes at a sufficiently large dissenter density.
	The intriguing phase behavior observed in simulations can be further accounted for by the coarse-grained continuous theory, provided that we take into consideration the higher-order orientation modes, rather than just density and polarity fields as usual~\cite{chate2020dry,toner2024physics}.
	
	While we have only studied the two-species VM with Voronoi neighbors in this work, given the generality of our coarse-grained continuous theory, we hope that our results could be generalized to other topological models such as the $k$NN-VM.
	Although due to fluctuations, traveling bands can already arise in the one-species $k$NN-VM near the flocking transition~\cite{martin2021fluctuation}, the presence of a small fraction of dissenters could further induce a new band regime at low noise strength far below the flocking transition, as in our Voronoi case.
	Actually, in the two-species Voronoi-VM with a low collision rate $\alpha_0=1/6$ studied in Appendix~\ref{app:model-low-rates}, one particle, on average, only ``collides'' with one of its Voronoi neighbors per time step, whose microscopic dynamics is already very similar to that in the $k$NN-VM with $k=1$.
	Given that the reentrant phase behavior also arises in the former case (Fig.~\ref{fig:f5} in Appendix~\ref{app:model-low-rates}), we expect the similar behavior could arise in the latter case and other $k$NN flocking models in the presence of dissenters.
	
	Also due to the generality of our coarse-grained continuous theory, in addition to the aligner-dissenter mixtures, our results should also hold for the case where two species are both aligners but suffer from different noise strength, such that one species is in the ordered phase while the other remains disordered.
	While temperature (noise) differences can induce interesting results in scalar active matter~\cite{grosberg2015nonequilibrium,weber2016binary,damman2024algebraic}, our results here imply that they could also play a nontrivial role in aligning active matter~\cite{chate2020dry} in the presence of topological interactions.
	
	Ultimately, our results suggest that, compared with their metric counterparts, topological flocks are much more sensitive to population heterogeneity such as weak nonreciprocity and temperature differences, which can induce the reentrant phase behavior and therefore play a qualitatively different role than noise.
        Given that population heterogeneity is unavoidable in natural flocking systems, we hope the rich phase behavior predicted by our minimal description could be observed in macroscopic living systems~\cite{ballerini2008interaction,Gautrais2012DecipheringII,camperi2012spatially,ginelli2015intermittent,pearce2014role,Jiang2017IdentifyingIN,jhawar2020noise,moussaid2011simple}, where topological interactions are shown to be dominated.
        On the other hand, population heterogeneity is likely necessary for robotic swarms to perform more complex tasks~\cite{dorigo2013swarmanoid}, in which case our results suggest that, when the constituting robots communicate through topological interactions~\cite{yasuda2014self}, the robustness of their homogeneous flocking state can be enhanced by appropriate noise instead.
	
	\acknowledgments
        We thank Beno\^{\i}t Mahault and Xiaqing Shi for useful comments.
	This work has received support from the Innovation Program for Quantum Science and Technology under Grant No.~2024ZD0300101, the National Natural Science Foundation of China under Grant Nos.~12174184, 12347102, as well as the Max Planck School Matter to Life and the MaxSynBio Consortium, which are jointly funded by the Federal Ministry of Education and Research (BMBF) of Germany, and the Max Planck Society.
	
	\appendix
	\section{Binary Voronoi-VM with low collision rates}\label{app:model-low-rates}
	
	Here we introduce a microscopic model, in which two neighboring particles collide at a rate $\alpha_0$ at every time step.
	To mimic binary collisions, it is convenient to use the continuous-time updating scheme, with which the aligning interactions become additive.
	In 2D, particle $i$ of species $S\in\{A, B\}$ is defined by its position $\mathbf{r}_{i,S}(t)$ and orientation $\theta_{i,S}(t)$, which evolve according to
	\begin{subequations} \label{eq:micro-model-low-alpha}
		\begin{align}
			\dot{\mathbf{r}}_{i, S}&= v_0 \hat{\mathbf{e}}(\theta_{i, S}), \\
			\dot{\theta}_{i, S} &= \sum_{\substack{j\sim i \\ X_{ij} \leq \alpha_0}} J_{SS'} \sin(\theta_{j, S'}-\theta_{i, S}) + \sqrt{2D}\xi_i(t),  \label{eq:micro-model-low-alpha-b}
		\end{align}
	\end{subequations}
	where overdots denote time derivative, 
	particles $j$ are the Voronoi neighbors of $i$, $J_{SS'}$ is the coupling between species $S$ and $S'$, $\xi_i$ is a white noise with zero mean and unit variance, while $D$ denotes the corresponding rotational diffusivity assumed equal for the two species.
	Here we still assume $J_{AA}=J_{AB}=1$ and $J_{BA}=J_{BB}=0$, such that $A$ particles are aligners and $B$ dissenters.
	
	In Eq.~\eqref{eq:micro-model-low-alpha-b}, we have introduced a random variable $X_{ij}$ uniformly distributed in $[0, 1]$. At every time step, for each pair of neighboring particles $i$ and $j$ identified by the Voronoi tessellation, a random $X_{ij}$ is generated, and particles $i$ and $j$ will ``collide'' with each other if $X_{ij}\leq \alpha_0$, otherwise they will be ``invisible'' to each other. 
	Then for $\alpha_0=1$, Eq.~\eqref{eq:micro-model-low-alpha} reduces to the conventional binary continuous-time VM~\cite{fruchart2021non} albeit with topological interactions here, while for $\alpha_0\ll 1$, binary collisions become dominated.
	
\begin{figure}
		\centering
		\includegraphics[width =\linewidth]{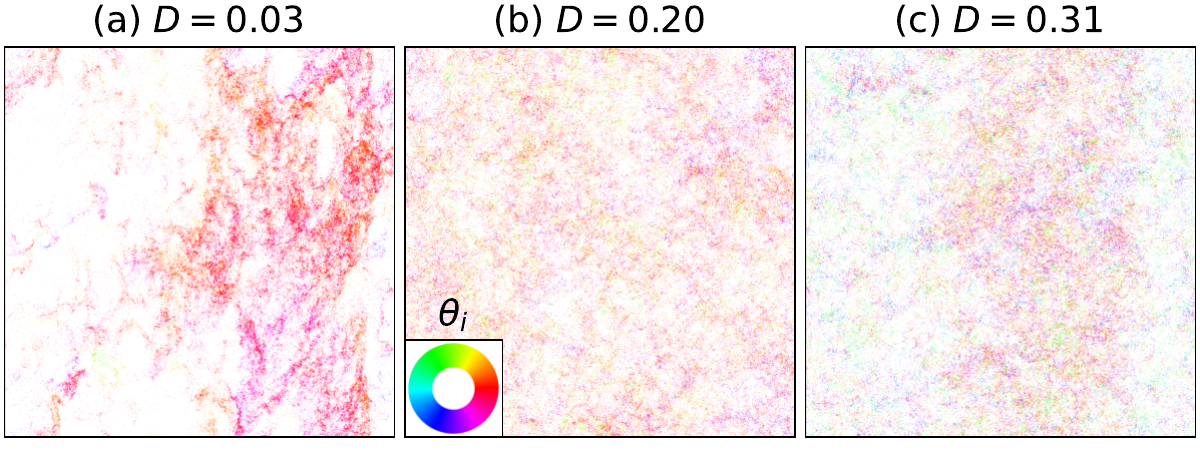}
		\caption{Reentrant phase behavior in the microscopic model~\eqref{eq:micro-model-low-alpha} with fixed $\bar{\rho}_B=0.03$ and varied noise strength $D$. Other parameters: total mean density $\rho_0=1$,  collision rate $\alpha_0=1/6$, system size $L=400$, time-step size $dt=0.1$.
		}
		\label{fig:f5}
	\end{figure}

	We set $\alpha_0=1/6$, so that on average one particle aligns only with one of its neighbors per time step.
	By simulations, we find that with increasing noise strength $D$, the reentrant phase behavior can arise for small $\bar{\rho}_B$, as shown in Fig.~\ref{fig:f5}.
	Hence, we believe that the microscopic models~\eqref{eq:micro-model} and~\eqref{eq:micro-model-low-alpha} should share qualitatively similar collective properties.

	\section{Derivation of continuous theory} \label{app:deriv-eq}
	Following the one-species case~\cite{peshkov2012continuous}, we derive nonlinear field equations for the two-species flocks with metric-free interactions.
	Instead of coarse gaining the microscopic model~\eqref{eq:micro-model} explicitly, our starting point is a simplified Vicsek-style model in which point-like aligners and dissenters move at a constant speed $v_0$ along their heading $\theta_i$ ballistically until they experience either a self-diffusion event, or a binary collision that changes the headings of the two particles~\cite{bertin2006boltzmann}.
    This simplified model shares the same microscopic dynamics as the microscopic model~\eqref{eq:micro-model-low-alpha} in the $\alpha_0 \ll 1$ limit. 
	The evolution of the one-particle phase-space distribution $f_S(\mathbf{r},\theta, t)$ for species $S$ is governed by the Boltzmann equation
	\begin{subequations} \label{eq:boltzmann}
		\begin{align}
			\partial_t f_A + v_0\hat{\mathbf{e}}(\theta)\cdot \nabla f_A &= I_\mathrm{dif}[f_A] + I_\mathrm{col}[f_A, f_B], \\
			\partial_t f_B + v_0\hat{\mathbf{e}}(\theta)\cdot \nabla f_B &= I_\mathrm{dif}[f_B],
		\end{align}
	\end{subequations}
	where $I_\mathrm{dif}[f_S]$ accounts for self-diffusion events which happen for the both species $S\in\{A, B\}$,
	while $I_{\mathrm{col}}[f_A,f_B]$ describes collision events which only apply to aligners from species $A$ but have no effects on dissenters from species $B$.
	
	In self-diffusion, $\theta_i$ is changed into $\theta'_i=\theta_i + \zeta$ with a probability $\lambda$ per unit time, where $\zeta$ is a random variable drawn from a symmetric distribution $P_\eta(\zeta)$ of variance $\eta^2$.
	For simplicity, we use a Gaussian distribution $P_\eta(\zeta)=\frac{1}{\eta\sqrt{2\pi}}\exp({-\zeta^2/2\eta^2})$.
	The corresponding self-diffusion integral is
	\begin{equation} \label{eq:I_dif}
		\begin{aligned}
			I_\mathrm{dif}[f_S] &= -\lambda f_S(\theta) +\lambda \int_{-\pi}^\pi \mathrm{d}\theta'\int\mathrm{d}\zeta P_\eta(\zeta) \\
			&\times \delta_{2\pi} (\theta'-\theta +\zeta) f_S(\theta'),
		\end{aligned}
	\end{equation}
	where $\delta_{2\pi}(\theta)=1$ for $\theta=2\pi m$ and zero otherwise, with $m$ an arbitrary integer.
	
	In collision events, we assume that particles from species $A$ collide with each Voronoi neighbor at rate $\alpha_0$ per unit time. 
	After one collision event, $\theta_i$ is changed to $\theta'_i=\Psi(\theta_i, \theta_j)+\zeta$, where $\theta_j$ is the heading of the chosen neighbor colliding with $\theta_i$, the noise $\zeta$ is also drawn from $P_\eta(\zeta)$ for simplicity, and $\Psi(\theta_1,\theta_2)\equiv \arg(e^{i\theta_1}+e^{i\theta_2})$ describes ferromagnetic alignment.
	In the small $\alpha_0$ limit, binary interactions dominate, taking place with a rate $\alpha= \mathcal{N}_i^t \alpha_0$, where number of neighbors $\mathcal{N}_i^t\approx 6$ for the Voronoi tessellation. Besides, in the small $\alpha_0$ limit, orientations are decorrelated between collisions.
	With these assumptions~\cite{peshkov2012continuous}, we can write
	\begin{equation}\label{eq:I_col}
		\begin{aligned}
			I_{\mathrm{col}}&[f_A, f_B] = -\alpha f_A(\theta) + \frac{\alpha}{\rho(\mathbf{r}, t)} \int_{-\pi}^\pi \mathrm{d}\theta_1\int_{-\pi}^\pi \mathrm{d}\theta_2 \\
			&\times \int_{-\infty}^\infty \mathrm{d}\zeta P_\eta(\zeta) f_A(\theta_1) [f_A(\theta_2)+f_B(\theta_2)] \\
			&\times \delta_{2\pi}(\Psi (\theta_1, \theta_2) -\theta +\zeta),
		\end{aligned}
	\end{equation}
	where $\rho(\mathbf{r}, t)$ denotes the total local density of the both species defined as
	\begin{equation}\label{eq:rho}
		\rho(\mathbf{r},t)=\rho_A(\mathbf{r}, t)+\rho_B({\mathbf{r}, t})=\int_{-\pi}^\pi \mathrm{d}\theta(f_A + f_B).
	\end{equation}
	Different from the metric case~\cite{bertin2006boltzmann}, here we follow Ref.~\cite{peshkov2012continuous} and assume the ``collision kernel'' is independent from relative angles and inversely proportional to the total local density $\rho(\mathbf{r}, t)$.
	In this way, Eq.~\eqref{eq:boltzmann}, together with the definitions of Eqs.~\eqref{eq:I_dif}-\eqref{eq:rho}, is left unchanged when we normalize $f_A$, $f_B$ and $\rho$ by an arbitrary factor simultaneously, in agreement with the basic properties of models with metric-free interactions.
	Without loss of generality, in the following we rescale time and space to set $\lambda=v_0=1$.
	
	To obtain the equations for the hydrodynamic fields, we expand $f_S(\mathbf{r},\theta, t)$ in angular Fourier series, yielding the angular Fourier modes $f_{k,S}(\mathbf{r}, t)=\int_{-\pi}^\pi \mathrm{d}\theta f_S(\mathbf{r},\theta, t)e^{ik\theta}$, with which the Boltzmann equation~\eqref{eq:boltzmann} yields the infinite hierarchy shown in Eq.~\eqref{eq:kinetic} of the main text, where $P_k=\int_{-\infty}^\infty \mathrm{d}\zeta P_\eta(\zeta)e^{ik\zeta}$ is the Fourier transform of $P_\eta$, while $I_{kl}$ is an integral depending on the alignment rule $\Psi$.
	For ferromagnetic alignment $\Psi(\theta_1,\theta_2)= \arg(e^{i\theta_1}+e^{i\theta_2})$, $I_{kl}$ reads 
	\begin{equation} \label{eq:I_kl}
		I_{kl} = \frac{1}{2\pi} \int_{-\pi}^\pi \mathrm{d}\theta \cos [(l-k/2)\theta].
	\end{equation}
	Besides, we need comment that the collision rate $\alpha$ presented in Eq.~\eqref{eq:kinetic} has been expressed in rescaled units, such that $\alpha$ is not necessarily smaller than $1$ in the units chosen, for which $v_0=\lambda=1$.
	
	The transport coefficients for the hydrodynamic equations~\eqref{eq:hydro} read
	\begin{equation} \label{eq:coefficients-hydro}
		\begin{aligned}
			\mu &= \left(
			\frac{2\alpha(2{\rho}_A+\bar{\rho}_B)}{\pi ({\rho}_A+\bar{\rho}_B)} + 1
			\right) P_1 - (1+\alpha), \\
			\nu &=\frac{1}{4(\alpha + 1 - P_2)}, \quad
			\gamma=\frac{4\alpha\nu}{\rho_A+\bar{\rho}_B}\left(
			P_2 - \frac{2P_1}{3\pi}
			\right), \\
			\kappa &= \frac{4\alpha\nu}{{\rho}_A+\bar{\rho}_B} \left(
			P_2 + \frac{2P_1}{3\pi}
			\right),
			\xi = \left(
			\frac{4\alpha}{\rho_A+\bar{\rho}_B}\right)^2 \frac{\nu}{3\pi} P_1 P_2.
		\end{aligned}
	\end{equation}
	When $\bar{\rho}_B=0$, we go back to the one-species case, and these coefficients are reduced to those in Ref.~\cite{peshkov2012continuous}.
	Given that $0 < P_k < 1$ for $k>0$, $\nu$, $\kappa$ and $\xi$ are always positive, while $\mu$ can change sign as mentioned in the main text.
	
	\section{Details of linear instability analysis} \label{app:linear-instability}
	
	\subsection{Hydrodynamic level}\label{app:linear-instability-hydro}
	
	Following Ref.~\cite{marchetti2013hydrodynamics}, we write $\mathbf{w}_A$ in terms of its magnitude and direction, letting $\mathbf{w}_A = w \hat{\mathbf{n}}$, where we have omitted the subscript $A$ for magnitude $w$ and unit vector $\hat{\mathbf{n}}$ in order to lighten the notation.
	Perturbations in $\mathbf{w}_A$ can then be written as $\delta \mathbf{w}_A= \hat{\mathbf{n}}_0 \delta w + w_0\delta \mathbf{n}$, where $w_0=\sqrt{\mu/\xi}$,
	and $\hat{\mathbf{n}}_0\cdot \delta \mathbf{n}=0$ to ensure $|\hat{\mathbf{n}}|=1$ to linear order.
	Setting $\hat{\mathbf{n}}_0=\hat{\mathbf{e}}_x$ to make the $x$ axis point along $\hat{\mathbf{n}}_0$, then in two dimensions $\delta \mathbf{n}=\delta n\hat{\mathbf{e}}_y$ and $\delta \mathbf{w}_A=\delta w \hat{\mathbf{e}}_x + w_0\delta n \hat{\mathbf{e}}_y$. 
	Substituting $\rho_A=\bar{\rho}_A + \delta \rho_A$ and $\mathbf{w}_A=w_0 \hat{\mathbf{n}}_0 + \delta \mathbf{w}_A$ into Eq.~\eqref{eq:hydro}, we obtained the linearized equations
	\begin{subequations} \label{eq:lin-hydro}
		\begin{align}
			\partial_t \delta \rho_A &= -(\partial_x \delta w + w_0 \partial_y \delta n), \label{eq:lin-hydro-a}\\
			\partial_t \delta w &=  (\mu'-\xi' w_0^2)w_0 \delta \rho_A -\gamma w_0 \partial_x \delta w -\frac{1}{2}\partial_x\delta \rho_A \notag\\
			&+\nu \nabla^2 \delta w -\kappa w_0^2 \partial_y \delta n -2\xi w_0^2 \delta w , \label{eq:lin-hydro-b}\\
			w_0\partial_t \delta n &= -\gamma w_0^2\partial_x\delta n -\frac{1}{2}\partial_y\delta \rho_A + w_0\nu \nabla^2 \delta n \notag \\
			&+\kappa w_0 \partial_y \delta w,
		\end{align}
	\end{subequations}
	where $\mu, \gamma, \kappa, \xi$ are evaluated at mean density $\bar{\rho}_A$ and primes indicate derivation with respect to $\rho_A$.
	Introducing a vector $\mathbf{u}(\mathbf{r}, t)=(\delta \rho_A(\mathbf{r}, t), \delta w(\mathbf{r},t), \delta n(\mathbf{r}, t))$
	and substituting the ansatz $\mathbf{u}(\mathbf{r},t)=\exp(\sigma t+i\mathbf{q}\cdot \mathbf{r})\mathbf{u}_\mathbf{q}(t)$ into Eq.~\eqref{eq:lin-hydro}, we find the evolution of fluctuations in Fourier space are dictated by
	\begin{equation}
		\partial_t\mathbf{u}_\mathbf{q} (t) = \underline{M}(\mathbf{q})\mathbf{u}_\mathbf{q}(t),
	\end{equation}
	where the matrix $\underline{M}$ reads
       \begin{equation} \label{eq:M}
			\underline{M}(\mathbf{q})=\begin{pmatrix}
				0 & -iq_x & -iq_y w_0\\
				M_{ w\rho} & M_{w w} & -iq_y\kappa w_0^2 \\
				-i\frac{q_y}{2}/w_0 & iq_y\kappa & -iq_x \gamma w_0 - q^2 \nu
			\end{pmatrix},
		\end{equation}
	where $q_x\equiv \mathbf{q}\cdot \hat{\mathbf{e}}_x$, $q_y\equiv \mathbf{q}\cdot \hat{\mathbf{e}}_y$, $q\equiv |\mathbf{q}|$,
    $M_{w\rho}(\mathbf{q})\equiv -\frac{iq_x}{2} + (\mu'-\xi'w_0^2)w_0$ and
    $M_{w w}(\mathbf{q})\equiv -iq_x\gamma w_0 -\nu q^2 -2\xi w_0^2$.

	\subsubsection{Near the order-disorder transition}\label{app:near-order-disorder}
	As the mean-field transition is approached from the ordered phase, i.e., for $\mu\to 0^+$ and $w_0\to 0^+$, the scaling ansatz~\eqref{eq:scaling-ansatz} suggests $\partial_t\sim\partial_{x,y}\sim w_0 \sim \delta \rho_A \sim \varepsilon$.
	Then Eq.~\eqref{eq:lin-hydro-a} suggests $\delta w\sim\varepsilon$ such that $\partial_t\delta \rho_A\sim\partial_x \delta w\sim \varepsilon^2$ while the remaining term $w_0\partial_y \delta n\sim \varepsilon^3$ is much smaller and thus can be ignored. Similarly, we can neglect the term in $\kappa$ in Eq.~\eqref{eq:lin-hydro-b} which is of higher order than the others (except for $-\xi' w_0^3 \delta \rho_A$ which we still retain, as it comes from the quartic part of the Ginzburg-Landau term).
	Then the dynamics of $\delta \rho_A$ and $\delta w$ are decoupled from $\delta n$, and thus Eq.~\eqref{eq:lin-hydro} is reduced to
	\begin{subequations} \label{eq:lin-hydro-rho-w}
		\begin{align}
			\partial_t \delta \rho_A &= -iq_x \delta w, \\
			\partial_t \delta w &= \left((\mu'-\xi' w_0^2)w_0 - \frac{i}{2} q_x\right)\delta \rho_A \notag \\
			&- (i\gamma w_0 q_x + \nu q^2 +2\xi w_0^2)\delta w.
		\end{align}
	\end{subequations}
	Solving the corresponding quadratic eigenvalue problem, we find the imaginary parts of growth rates $\sigma$ are always nonzero, while their largest real part reads
	\begin{equation}\label{eq:sigma_RE}
		\Re(\sigma) = \frac{1}{2}\left(
		\sqrt{\frac{a+\sqrt{a^2+ 4 b^2}}{2}}
		- (\nu q^2 + 2\xi w_0^2)
		\right),
	\end{equation}
	where $a\equiv (\nu q^2 + 2\xi w_0^2)^2 - (\gamma^2w_0^2 +2)q_x^2$ and
	$b \equiv q_x w_0(\gamma(\nu q^2 + 2\xi w_0^2)-2(\mu'-\xi' w_0^2))$.
	After some algebra, we find $\Re(\sigma)>0$ iff
	\begin{equation}
		\begin{aligned}
			w_0^2 (2(\mu'-\xi' w_0^2) - \gamma (\nu q^2 + 2\xi w_0^2))^2   \\>(\nu q^2 + 2\xi w_0^2)^2 (2+w_0^2 \gamma^2),
		\end{aligned}
	\end{equation}
	which is reduced to Eq.~\eqref{eq:instab-condi-1D} as $q\to0$ by noting that $w_0^2=\mu/\xi$.
	
	\subsubsection{Numerical method for the cubic eigenvalue problem}\label{app:num-3rd}
	To calculate the growth rates of linear system~\eqref{eq:lin-hydro-M}, for given $\bar{\rho}_B$, $\eta$ and wave vector $\mathbf{q}=q(\cos\phi, \sin\phi)$, from Eqs.~\eqref{eq:coefficients-hydro} and~\eqref{eq:M} we obtain the corresponding matrix $\underline{M}(\mathbf{q})$, whose eigenvalues $\sigma$ can be calculated numerically. 
	To detect long-wave instability, we only need to check the sign of $\Re(\sigma)$ at a small wavenumber $q^*$. 
	For given $\bar{\rho}_B$, $\eta$, and fixed $q=q^*$, $\Re(\sigma)$ will change with varied $\phi$.
	We denote by $\phi_m$ the most unstable direction along which $\Re(\sigma)$ gets its maximal value.
	Instability will arise at given $(\bar{\rho}_B, \eta)$ if this maximal $\Re(\sigma)$ is positive.
	Scanning the $(\bar{\rho}_B,\eta)$ plane, we then get the linear stability diagram shown in Fig.~\ref{fig:f4}(a), for which we have used $q^*=10^{-4}$.
	We have checked that the results remain nearly unchanged for other $q^*$ provided that it is not too large.
	
	\subsection{Kinetic level}\label{app:linear-instability-kinetic}
	Here we show the details how we perform linear instability analysis at kinetic level leading to the linear diagrams shown in Fig.~\ref{fig:f4}(b-d).
	Given that the stable solution for species $B$ has been given by Eq.~\eqref{eq:solution-f_k_B}, we only need to perform the linear instability analysis on the kinetic equation for species $A$.
	To get closed equations for species $A$, we truncate the infinite  hierarchy~\eqref{eq:kinetic-a} at a given order $K$, i.e., setting $f_{k,A}=0$ for all $|k|>K$.
	At kinetic level $K$,
	the homogeneous solution $\{\bar{f}_{k, A}\}$ with $k=0, 1, \dots, K$ satisfies
	\begin{equation} \label{eq:kinitic-homo-sol}
		\begin{aligned}
			(P_k-1-\alpha) \bar{f}_{k,A} + \frac{\alpha}{\rho_0} P_k \sum_{\substack{l=-K\\ |k-l|\leq K}}^{K} I_{kl} \bar{f}_{k-l, A} (\bar{f}_{l, A}+ \delta_{0,l}\bar{\rho}_B)
		\end{aligned}
	\end{equation}
	for any $ 0\leq k\leq K$,
	where $P_k(\eta)=\exp(-k^2\eta^2/2)$ and $I_{kl}$ is given by Eq.~\eqref{eq:I_kl}, while $\delta_{m,n}=1$ if $m=n$ and zero otherwise.
	Due to rotational invariance, we choose the global polar order $\bar{f}_{1,A}$ to be real, thus the other modes of the corresponding solution are all real by symmetry.
	Then solution $\{\bar{f}_{k, A}\}$ can be found by solving Eq.~\eqref{eq:kinitic-homo-sol} numerically.
	
	To examine the stability of of the numerical solution $\{\bar{f}_{k, A}\}$, we linearize the modes as $f_{k, A}=\bar{f}_{k, A}+\delta f_{k, A}$. The corresponding equations for the perturbations read
	\begin{equation}\label{eq:lin-kinitic}
		\begin{aligned}
			\partial_t \delta f_{k, A} &=- \frac{1}{2}(\triangledown^* \delta f_{k+1, A} +\triangledown \delta f_{k-1, A}) + \frac{\alpha}{\rho_0} P_k I_{k, 0} \bar{\rho}_B \delta f_{k, A}\\ 
			&+ (P_k-1-\alpha) (\delta f_{k, A} + \frac{\bar{f}_{k, A}}{\rho_0} \delta \rho) \\
			&+ \frac{\alpha}{\rho_0} P_k \sum_{\substack{l=-K\\ |k-l|\leq K}}^K (I_{k, l}+ I_{k, k-l}) \bar{f}_{k-l, A} \delta f_{l, A}.
		\end{aligned}
	\end{equation}
	Given that $\delta f_{k, A}$ are complex and $\delta f_{k, A}=\delta f_{-k, A}^*$, we can define $\delta f_{k, A}= g_k + ih_k$ with real $g_k=g_{-k}$ and $h_{k}=-h_{-k}$.
	From Eq.~\eqref{eq:lin-kinitic}, we find that $\{g_k\}$ evolve according to
	\begin{equation}\label{eq:lin-kinitic-g}
		\begin{aligned}
			\partial_t g_k &=-\frac{1}{2} \left[\partial_x (g_{k+1}+g_{k-1}) + \partial_y (h_{k+1}-h_{k-1})\right] \\
			&+\frac{\alpha}{\rho_0}P_k I_{k, 0}\bar{\rho}_B  g_k + (P_k-1-\alpha) (g_k + \frac{\bar{f}_{k, A}}{\rho_0}g_0) 
			\\
			&+\frac{\alpha}{\rho_0} P_k \sum_{\substack{l=-K\\ |k-l|\leq K}}^K (I_{k, l}+ I_{k, k-l})\bar{f}_{k-l, A}g_l,
		\end{aligned}
	\end{equation}
	while $\{h_k\}$ evolve according to
	\begin{equation}\label{eq:lin-kinitic-h}
		\begin{aligned}
			\partial_t h_k&=\frac{1}{2}[\partial_y (g_{k+1}-g_{k-1})
			-\partial_x (h_{k-1}+h_{k+1})]  \\
			&+\frac{\alpha}{\rho_0}P_k I_{k, 0}\bar{\rho}_B  h_k+ (P_k-1-\alpha)(h_k + \frac{\bar{f}_{k, A}}{\rho_0}h_0)  \\
			&+\frac{\alpha}{\rho_0} P_k \sum_{\substack{l=-K\\ |k-l|\leq K}}^K (I_{k, l}+ I_{k, k-l})\bar{f}_{k-l, A}h_l.
		\end{aligned}
	\end{equation}
	
	Defining $\mathbf{g}=(g_0, g_1,\dots, g_K)$ and $\mathbf{h}=(h_0, h_1,\dots, h_K)$, Eqs.~\eqref{eq:lin-kinitic-g} and~\eqref{eq:lin-kinitic-h} can be written as after the spatial Fourier transform, 
	\begin{equation}\label{eq:lin-kinitic-g-h-q}
		\partial_t\begin{pmatrix}
			\mathbf{g}_\mathbf{q} \\
			\mathbf{h}_\mathbf{q}
		\end{pmatrix}
		= \begin{pmatrix}
			\underline{M}_{gg}(\mathbf{q}) & \underline{M}_{gh}(\mathbf{q}) \\
			\underline{M}_{hg}(\mathbf{q}) & \underline{M}_{hh}(\mathbf{q})
		\end{pmatrix}
		\begin{pmatrix}
			\mathbf{g}_\mathbf{q}\\
			\mathbf{h}_\mathbf{q}
		\end{pmatrix},
	\end{equation}
	where matrices $\underline{M}_{XY}$ for $X, Y \in\{g, h\}$ can be obtained from Eqs.~\eqref{eq:lin-kinitic-g} and~\eqref{eq:lin-kinitic-h} straightforwardly by replacing $\partial_x$ and $\partial_y$ with $iq_x$ and $iq_y$ respectively. 
	Then we can solve the eigenvalue problem~\eqref{eq:lin-kinitic-g-h-q} numerically to know the stability of the solution $\{\bar{f}_{k, A}\}$.

	\bibliography{ref.bib}

\begin{thebibliography}{72}%
\makeatletter
\providecommand \@ifxundefined [1]{%
 \@ifx{#1\undefined}
}%
\providecommand \@ifnum [1]{%
 \ifnum #1\expandafter \@firstoftwo
 \else \expandafter \@secondoftwo
 \fi
}%
\providecommand \@ifx [1]{%
 \ifx #1\expandafter \@firstoftwo
 \else \expandafter \@secondoftwo
 \fi
}%
\providecommand \natexlab [1]{#1}%
\providecommand \enquote  [1]{``#1''}%
\providecommand \bibnamefont  [1]{#1}%
\providecommand \bibfnamefont [1]{#1}%
\providecommand \citenamefont [1]{#1}%
\providecommand \href@noop [0]{\@secondoftwo}%
\providecommand \href [0]{\begingroup \@sanitize@url \@href}%
\providecommand \@href[1]{\@@startlink{#1}\@@href}%
\providecommand \@@href[1]{\endgroup#1\@@endlink}%
\providecommand \@sanitize@url [0]{\catcode `\\12\catcode `\$12\catcode `\&12\catcode `\#12\catcode `\^12\catcode `\_12\catcode `\%12\relax}%
\providecommand \@@startlink[1]{}%
\providecommand \@@endlink[0]{}%
\providecommand \url  [0]{\begingroup\@sanitize@url \@url }%
\providecommand \@url [1]{\endgroup\@href {#1}{\urlprefix }}%
\providecommand \urlprefix  [0]{URL }%
\providecommand \Eprint [0]{\href }%
\providecommand \doibase [0]{https://doi.org/}%
\providecommand \selectlanguage [0]{\@gobble}%
\providecommand \bibinfo  [0]{\@secondoftwo}%
\providecommand \bibfield  [0]{\@secondoftwo}%
\providecommand \translation [1]{[#1]}%
\providecommand \BibitemOpen [0]{}%
\providecommand \bibitemStop [0]{}%
\providecommand \bibitemNoStop [0]{.\EOS\space}%
\providecommand \EOS [0]{\spacefactor3000\relax}%
\providecommand \BibitemShut  [1]{\csname bibitem#1\endcsname}%
\let\auto@bib@innerbib\@empty
\bibitem [{\citenamefont {Vicsek}\ \emph {et~al.}(1995)\citenamefont {Vicsek}, \citenamefont {Czir{\'o}k}, \citenamefont {Ben-Jacob}, \citenamefont {Cohen},\ and\ \citenamefont {Shochet}}]{vicsek1995novel}%
  \BibitemOpen
  \bibfield  {author} {\bibinfo {author} {\bibfnamefont {T.}~\bibnamefont {Vicsek}}, \bibinfo {author} {\bibfnamefont {A.}~\bibnamefont {Czir{\'o}k}}, \bibinfo {author} {\bibfnamefont {E.}~\bibnamefont {Ben-Jacob}}, \bibinfo {author} {\bibfnamefont {I.}~\bibnamefont {Cohen}},\ and\ \bibinfo {author} {\bibfnamefont {O.}~\bibnamefont {Shochet}},\ }\bibfield  {title} {\bibinfo {title} {Novel type of phase transition in a system of self-driven particles},\ }\href {https://doi.org/https://doi.org/10.1103/PhysRevLett.75.1226} {\bibfield  {journal} {\bibinfo  {journal} {Phys. Rev. Lett.}\ }\textbf {\bibinfo {volume} {75}},\ \bibinfo {pages} {1226} (\bibinfo {year} {1995})}\BibitemShut {NoStop}%
\bibitem [{\citenamefont {Chat{\'e}}(2020)}]{chate2020dry}%
  \BibitemOpen
  \bibfield  {author} {\bibinfo {author} {\bibfnamefont {H.}~\bibnamefont {Chat{\'e}}},\ }\bibfield  {title} {\bibinfo {title} {Dry aligning dilute active matter},\ }\href {https://doi.org/https://doi.org/10.1146/annurev-conmatphys-031119-050752} {\bibfield  {journal} {\bibinfo  {journal} {Annual Review of Condensed Matter Physics}\ }\textbf {\bibinfo {volume} {11}},\ \bibinfo {pages} {189} (\bibinfo {year} {2020})}\BibitemShut {NoStop}%
\bibitem [{\citenamefont {Toner}(2024)}]{toner2024physics}%
  \BibitemOpen
  \bibfield  {author} {\bibinfo {author} {\bibfnamefont {J.}~\bibnamefont {Toner}},\ }\href {https://doi.org/https://doi.org/10.1017/9781108993623} {\emph {\bibinfo {title} {The Physics of Flocking: Birth, Death, and Flight in Active Matter}}}\ (\bibinfo  {publisher} {Cambridge University Press},\ \bibinfo {year} {2024})\BibitemShut {NoStop}%
\bibitem [{\citenamefont {Solon}(2024)}]{solon2024thirty}%
  \BibitemOpen
  \bibfield  {author} {\bibinfo {author} {\bibfnamefont {A.}~\bibnamefont {Solon}},\ }\bibfield  {title} {\bibinfo {title} {Thirty years of surprises about collective motion},\ }\href {https://doi.org/https://doi.org/10.1051/epn/2024309} {\bibfield  {journal} {\bibinfo  {journal} {Europhysics News}\ }\textbf {\bibinfo {volume} {55}},\ \bibinfo {pages} {28} (\bibinfo {year} {2024})}\BibitemShut {NoStop}%
\bibitem [{\citenamefont {Bain}\ and\ \citenamefont {Bartolo}(2019)}]{bain2019dynamic}%
  \BibitemOpen
  \bibfield  {author} {\bibinfo {author} {\bibfnamefont {N.}~\bibnamefont {Bain}}\ and\ \bibinfo {author} {\bibfnamefont {D.}~\bibnamefont {Bartolo}},\ }\bibfield  {title} {\bibinfo {title} {Dynamic response and hydrodynamics of polarized crowds},\ }\href {https://doi.org/10.1126/science.aat9891} {\bibfield  {journal} {\bibinfo  {journal} {Science}\ }\textbf {\bibinfo {volume} {363}},\ \bibinfo {pages} {46} (\bibinfo {year} {2019})}\BibitemShut {NoStop}%
\bibitem [{\citenamefont {Ballerini}\ \emph {et~al.}(2008)\citenamefont {Ballerini}, \citenamefont {Cabibbo}, \citenamefont {Candelier}, \citenamefont {Cavagna}, \citenamefont {Cisbani}, \citenamefont {Giardina}, \citenamefont {Lecomte}, \citenamefont {Orlandi}, \citenamefont {Parisi}, \citenamefont {Procaccini} \emph {et~al.}}]{ballerini2008interaction}%
  \BibitemOpen
  \bibfield  {author} {\bibinfo {author} {\bibfnamefont {M.}~\bibnamefont {Ballerini}}, \bibinfo {author} {\bibfnamefont {N.}~\bibnamefont {Cabibbo}}, \bibinfo {author} {\bibfnamefont {R.}~\bibnamefont {Candelier}}, \bibinfo {author} {\bibfnamefont {A.}~\bibnamefont {Cavagna}}, \bibinfo {author} {\bibfnamefont {E.}~\bibnamefont {Cisbani}}, \bibinfo {author} {\bibfnamefont {I.}~\bibnamefont {Giardina}}, \bibinfo {author} {\bibfnamefont {V.}~\bibnamefont {Lecomte}}, \bibinfo {author} {\bibfnamefont {A.}~\bibnamefont {Orlandi}}, \bibinfo {author} {\bibfnamefont {G.}~\bibnamefont {Parisi}}, \bibinfo {author} {\bibfnamefont {A.}~\bibnamefont {Procaccini}}, \emph {et~al.},\ }\bibfield  {title} {\bibinfo {title} {Interaction ruling animal collective behavior depends on topological rather than metric distance: Evidence from a field study},\ }\href {https://doi.org/https://doi.org/10.1073/pnas.0711437105} {\bibfield  {journal} {\bibinfo  {journal} {Proceedings of the national academy of sciences}\ }\textbf {\bibinfo
  {volume} {105}},\ \bibinfo {pages} {1232} (\bibinfo {year} {2008})}\BibitemShut {NoStop}%
\bibitem [{\citenamefont {Cavagna}\ \emph {et~al.}(2010)\citenamefont {Cavagna}, \citenamefont {Cimarelli}, \citenamefont {Giardina}, \citenamefont {Parisi}, \citenamefont {Santagati}, \citenamefont {Stefanini},\ and\ \citenamefont {Viale}}]{cavagna2010scale}%
  \BibitemOpen
  \bibfield  {author} {\bibinfo {author} {\bibfnamefont {A.}~\bibnamefont {Cavagna}}, \bibinfo {author} {\bibfnamefont {A.}~\bibnamefont {Cimarelli}}, \bibinfo {author} {\bibfnamefont {I.}~\bibnamefont {Giardina}}, \bibinfo {author} {\bibfnamefont {G.}~\bibnamefont {Parisi}}, \bibinfo {author} {\bibfnamefont {R.}~\bibnamefont {Santagati}}, \bibinfo {author} {\bibfnamefont {F.}~\bibnamefont {Stefanini}},\ and\ \bibinfo {author} {\bibfnamefont {M.}~\bibnamefont {Viale}},\ }\bibfield  {title} {\bibinfo {title} {Scale-free correlations in starling flocks},\ }\href {https://doi.org/https://doi.org/10.1073/pnas.1005766107} {\bibfield  {journal} {\bibinfo  {journal} {Proceedings of the National Academy of Sciences}\ }\textbf {\bibinfo {volume} {107}},\ \bibinfo {pages} {11865} (\bibinfo {year} {2010})}\BibitemShut {NoStop}%
\bibitem [{\citenamefont {Nishiguchi}\ \emph {et~al.}(2017)\citenamefont {Nishiguchi}, \citenamefont {Nagai}, \citenamefont {Chat{\'e}},\ and\ \citenamefont {Sano}}]{nishiguchi2017long}%
  \BibitemOpen
  \bibfield  {author} {\bibinfo {author} {\bibfnamefont {D.}~\bibnamefont {Nishiguchi}}, \bibinfo {author} {\bibfnamefont {K.~H.}\ \bibnamefont {Nagai}}, \bibinfo {author} {\bibfnamefont {H.}~\bibnamefont {Chat{\'e}}},\ and\ \bibinfo {author} {\bibfnamefont {M.}~\bibnamefont {Sano}},\ }\bibfield  {title} {\bibinfo {title} {Long-range nematic order and anomalous fluctuations in suspensions of swimming filamentous bacteria},\ }\href {https://doi.org/https://doi.org/10.1103/PhysRevE.95.020601} {\bibfield  {journal} {\bibinfo  {journal} {Physical Review E}\ }\textbf {\bibinfo {volume} {95}},\ \bibinfo {pages} {020601} (\bibinfo {year} {2017})}\BibitemShut {NoStop}%
\bibitem [{\citenamefont {Schaller}\ \emph {et~al.}(2010)\citenamefont {Schaller}, \citenamefont {Weber}, \citenamefont {Semmrich}, \citenamefont {Frey},\ and\ \citenamefont {Bausch}}]{schaller2010polar}%
  \BibitemOpen
  \bibfield  {author} {\bibinfo {author} {\bibfnamefont {V.}~\bibnamefont {Schaller}}, \bibinfo {author} {\bibfnamefont {C.}~\bibnamefont {Weber}}, \bibinfo {author} {\bibfnamefont {C.}~\bibnamefont {Semmrich}}, \bibinfo {author} {\bibfnamefont {E.}~\bibnamefont {Frey}},\ and\ \bibinfo {author} {\bibfnamefont {A.~R.}\ \bibnamefont {Bausch}},\ }\bibfield  {title} {\bibinfo {title} {Polar patterns of driven filaments},\ }\href {https://doi.org/https://doi.org/10.1038/nature09312} {\bibfield  {journal} {\bibinfo  {journal} {Nature}\ }\textbf {\bibinfo {volume} {467}},\ \bibinfo {pages} {73} (\bibinfo {year} {2010})}\BibitemShut {NoStop}%
\bibitem [{\citenamefont {Bricard}\ \emph {et~al.}(2013)\citenamefont {Bricard}, \citenamefont {Caussin}, \citenamefont {Desreumaux}, \citenamefont {Dauchot},\ and\ \citenamefont {Bartolo}}]{bricard2013emergence}%
  \BibitemOpen
  \bibfield  {author} {\bibinfo {author} {\bibfnamefont {A.}~\bibnamefont {Bricard}}, \bibinfo {author} {\bibfnamefont {J.-B.}\ \bibnamefont {Caussin}}, \bibinfo {author} {\bibfnamefont {N.}~\bibnamefont {Desreumaux}}, \bibinfo {author} {\bibfnamefont {O.}~\bibnamefont {Dauchot}},\ and\ \bibinfo {author} {\bibfnamefont {D.}~\bibnamefont {Bartolo}},\ }\bibfield  {title} {\bibinfo {title} {Emergence of macroscopic directed motion in populations of motile colloids},\ }\href {https://doi.org/https://doi.org/10.1038/nature12673} {\bibfield  {journal} {\bibinfo  {journal} {Nature}\ }\textbf {\bibinfo {volume} {503}},\ \bibinfo {pages} {95} (\bibinfo {year} {2013})}\BibitemShut {NoStop}%
\bibitem [{\citenamefont {Das}\ \emph {et~al.}(2024)\citenamefont {Das}, \citenamefont {Ciarchi}, \citenamefont {Zhou}, \citenamefont {Yan}, \citenamefont {Zhang},\ and\ \citenamefont {Alert}}]{das2024flocking}%
  \BibitemOpen
  \bibfield  {author} {\bibinfo {author} {\bibfnamefont {S.}~\bibnamefont {Das}}, \bibinfo {author} {\bibfnamefont {M.}~\bibnamefont {Ciarchi}}, \bibinfo {author} {\bibfnamefont {Z.}~\bibnamefont {Zhou}}, \bibinfo {author} {\bibfnamefont {J.}~\bibnamefont {Yan}}, \bibinfo {author} {\bibfnamefont {J.}~\bibnamefont {Zhang}},\ and\ \bibinfo {author} {\bibfnamefont {R.}~\bibnamefont {Alert}},\ }\bibfield  {title} {\bibinfo {title} {Flocking by turning away},\ }\href {https://doi.org/https://doi.org/10.1103/PhysRevX.14.031008} {\bibfield  {journal} {\bibinfo  {journal} {Physical Review X}\ }\textbf {\bibinfo {volume} {14}},\ \bibinfo {pages} {031008} (\bibinfo {year} {2024})}\BibitemShut {NoStop}%
\bibitem [{\citenamefont {Deseigne}\ \emph {et~al.}(2010)\citenamefont {Deseigne}, \citenamefont {Dauchot},\ and\ \citenamefont {Chat{\'e}}}]{deseigne2010collective}%
  \BibitemOpen
  \bibfield  {author} {\bibinfo {author} {\bibfnamefont {J.}~\bibnamefont {Deseigne}}, \bibinfo {author} {\bibfnamefont {O.}~\bibnamefont {Dauchot}},\ and\ \bibinfo {author} {\bibfnamefont {H.}~\bibnamefont {Chat{\'e}}},\ }\bibfield  {title} {\bibinfo {title} {Collective motion of vibrated polar disks},\ }\href {https://doi.org/https://doi.org/10.1103/PhysRevLett.105.098001} {\bibfield  {journal} {\bibinfo  {journal} {Physical review letters}\ }\textbf {\bibinfo {volume} {105}},\ \bibinfo {pages} {098001} (\bibinfo {year} {2010})}\BibitemShut {NoStop}%
\bibitem [{\citenamefont {Kumar}\ \emph {et~al.}(2014)\citenamefont {Kumar}, \citenamefont {Soni}, \citenamefont {Ramaswamy},\ and\ \citenamefont {Sood}}]{kumar2014flocking}%
  \BibitemOpen
  \bibfield  {author} {\bibinfo {author} {\bibfnamefont {N.}~\bibnamefont {Kumar}}, \bibinfo {author} {\bibfnamefont {H.}~\bibnamefont {Soni}}, \bibinfo {author} {\bibfnamefont {S.}~\bibnamefont {Ramaswamy}},\ and\ \bibinfo {author} {\bibfnamefont {A.}~\bibnamefont {Sood}},\ }\bibfield  {title} {\bibinfo {title} {Flocking at a distance in active granular matter},\ }\href {https://doi.org/https://doi.org/10.1038/ncomms5688} {\bibfield  {journal} {\bibinfo  {journal} {Nature communications}\ }\textbf {\bibinfo {volume} {5}},\ \bibinfo {pages} {4688} (\bibinfo {year} {2014})}\BibitemShut {NoStop}%
\bibitem [{\citenamefont {V{\'a}s{\'a}rhelyi}\ \emph {et~al.}(2018)\citenamefont {V{\'a}s{\'a}rhelyi}, \citenamefont {Vir{\'a}gh}, \citenamefont {Somorjai}, \citenamefont {Nepusz}, \citenamefont {Eiben},\ and\ \citenamefont {Vicsek}}]{vasarhelyi2018optimized}%
  \BibitemOpen
  \bibfield  {author} {\bibinfo {author} {\bibfnamefont {G.}~\bibnamefont {V{\'a}s{\'a}rhelyi}}, \bibinfo {author} {\bibfnamefont {C.}~\bibnamefont {Vir{\'a}gh}}, \bibinfo {author} {\bibfnamefont {G.}~\bibnamefont {Somorjai}}, \bibinfo {author} {\bibfnamefont {T.}~\bibnamefont {Nepusz}}, \bibinfo {author} {\bibfnamefont {A.~E.}\ \bibnamefont {Eiben}},\ and\ \bibinfo {author} {\bibfnamefont {T.}~\bibnamefont {Vicsek}},\ }\bibfield  {title} {\bibinfo {title} {Optimized flocking of autonomous drones in confined environments},\ }\href {https://doi.org/10.1126/scirobotics.aat3536} {\bibfield  {journal} {\bibinfo  {journal} {Science Robotics}\ }\textbf {\bibinfo {volume} {3}},\ \bibinfo {pages} {eaat3536} (\bibinfo {year} {2018})}\BibitemShut {NoStop}%
\bibitem [{\citenamefont {Tu}\ \emph {et~al.}(1998)\citenamefont {Tu}, \citenamefont {Toner},\ and\ \citenamefont {Ulm}}]{tu1998sound}%
  \BibitemOpen
  \bibfield  {author} {\bibinfo {author} {\bibfnamefont {Y.}~\bibnamefont {Tu}}, \bibinfo {author} {\bibfnamefont {J.}~\bibnamefont {Toner}},\ and\ \bibinfo {author} {\bibfnamefont {M.}~\bibnamefont {Ulm}},\ }\bibfield  {title} {\bibinfo {title} {Sound waves and the absence of galilean invariance in flocks},\ }\href {https://doi.org/https://doi.org/10.1103/PhysRevLett.80.4819} {\bibfield  {journal} {\bibinfo  {journal} {Physical review letters}\ }\textbf {\bibinfo {volume} {80}},\ \bibinfo {pages} {4819} (\bibinfo {year} {1998})}\BibitemShut {NoStop}%
\bibitem [{\citenamefont {Geyer}\ \emph {et~al.}(2018)\citenamefont {Geyer}, \citenamefont {Morin},\ and\ \citenamefont {Bartolo}}]{geyer2018sounds}%
  \BibitemOpen
  \bibfield  {author} {\bibinfo {author} {\bibfnamefont {D.}~\bibnamefont {Geyer}}, \bibinfo {author} {\bibfnamefont {A.}~\bibnamefont {Morin}},\ and\ \bibinfo {author} {\bibfnamefont {D.}~\bibnamefont {Bartolo}},\ }\bibfield  {title} {\bibinfo {title} {Sounds and hydrodynamics of polar active fluids},\ }\href {https://doi.org/https://doi.org/10.1038/s41563-018-0123-4} {\bibfield  {journal} {\bibinfo  {journal} {Nature materials}\ }\textbf {\bibinfo {volume} {17}},\ \bibinfo {pages} {789} (\bibinfo {year} {2018})}\BibitemShut {NoStop}%
\bibitem [{\citenamefont {Mahault}\ \emph {et~al.}(2019)\citenamefont {Mahault}, \citenamefont {Ginelli},\ and\ \citenamefont {Chat{\'e}}}]{mahault2019quantitative}%
  \BibitemOpen
  \bibfield  {author} {\bibinfo {author} {\bibfnamefont {B.}~\bibnamefont {Mahault}}, \bibinfo {author} {\bibfnamefont {F.}~\bibnamefont {Ginelli}},\ and\ \bibinfo {author} {\bibfnamefont {H.}~\bibnamefont {Chat{\'e}}},\ }\bibfield  {title} {\bibinfo {title} {{Quantitative Assessment of the Toner and Tu Theory of Polar Flocks}},\ }\href {https://doi.org/https://doi.org/10.1103/PhysRevLett.123.218001} {\bibfield  {journal} {\bibinfo  {journal} {Phys. Rev. Lett.}\ }\textbf {\bibinfo {volume} {123}},\ \bibinfo {pages} {218001} (\bibinfo {year} {2019})}\BibitemShut {NoStop}%
\bibitem [{\citenamefont {Tasaki}(2020)}]{tasaki2020hohenberg}%
  \BibitemOpen
  \bibfield  {author} {\bibinfo {author} {\bibfnamefont {H.}~\bibnamefont {Tasaki}},\ }\bibfield  {title} {\bibinfo {title} {Hohenberg-mermin-wagner-type theorems for equilibrium models of flocking},\ }\href {https://doi.org/https://doi.org/10.1103/PhysRevLett.125.220601} {\bibfield  {journal} {\bibinfo  {journal} {Physical Review Letters}\ }\textbf {\bibinfo {volume} {125}},\ \bibinfo {pages} {220601} (\bibinfo {year} {2020})}\BibitemShut {NoStop}%
\bibitem [{\citenamefont {Mahault}\ and\ \citenamefont {Chat{\'e}}(2021)}]{mahault2021long}%
  \BibitemOpen
  \bibfield  {author} {\bibinfo {author} {\bibfnamefont {B.}~\bibnamefont {Mahault}}\ and\ \bibinfo {author} {\bibfnamefont {H.}~\bibnamefont {Chat{\'e}}},\ }\bibfield  {title} {\bibinfo {title} {Long-range nematic order in two-dimensional active matter},\ }\href {https://doi.org/https://doi.org/10.1103/PhysRevLett.127.048003} {\bibfield  {journal} {\bibinfo  {journal} {Physical Review Letters}\ }\textbf {\bibinfo {volume} {127}},\ \bibinfo {pages} {048003} (\bibinfo {year} {2021})}\BibitemShut {NoStop}%
\bibitem [{\citenamefont {Fava}\ \emph {et~al.}(2024)\citenamefont {Fava}, \citenamefont {Gambassi},\ and\ \citenamefont {Ginelli}}]{fava2024strong}%
  \BibitemOpen
  \bibfield  {author} {\bibinfo {author} {\bibfnamefont {G.}~\bibnamefont {Fava}}, \bibinfo {author} {\bibfnamefont {A.}~\bibnamefont {Gambassi}},\ and\ \bibinfo {author} {\bibfnamefont {F.}~\bibnamefont {Ginelli}},\ }\bibfield  {title} {\bibinfo {title} {Strong casimir-like forces in flocking active matter},\ }\href {https://doi.org/https://doi.org/10.1103/PhysRevLett.133.148301} {\bibfield  {journal} {\bibinfo  {journal} {Physical Review Letters}\ }\textbf {\bibinfo {volume} {133}},\ \bibinfo {pages} {148301} (\bibinfo {year} {2024})}\BibitemShut {NoStop}%
\bibitem [{\citenamefont {Ginelli}\ and\ \citenamefont {Chat{\'e}}(2010)}]{ginelli2010relevance}%
  \BibitemOpen
  \bibfield  {author} {\bibinfo {author} {\bibfnamefont {F.}~\bibnamefont {Ginelli}}\ and\ \bibinfo {author} {\bibfnamefont {H.}~\bibnamefont {Chat{\'e}}},\ }\bibfield  {title} {\bibinfo {title} {Relevance of metric-free interactions in flocking phenomena},\ }\href {https://doi.org/https://doi.org/10.1103/PhysRevLett.105.168103} {\bibfield  {journal} {\bibinfo  {journal} {Physical review letters}\ }\textbf {\bibinfo {volume} {105}},\ \bibinfo {pages} {168103} (\bibinfo {year} {2010})}\BibitemShut {NoStop}%
\bibitem [{\citenamefont {Martin}\ \emph {et~al.}(2021)\citenamefont {Martin}, \citenamefont {Chat{\'e}}, \citenamefont {Nardini}, \citenamefont {Solon}, \citenamefont {Tailleur},\ and\ \citenamefont {Van~Wijland}}]{martin2021fluctuation}%
  \BibitemOpen
  \bibfield  {author} {\bibinfo {author} {\bibfnamefont {D.}~\bibnamefont {Martin}}, \bibinfo {author} {\bibfnamefont {H.}~\bibnamefont {Chat{\'e}}}, \bibinfo {author} {\bibfnamefont {C.}~\bibnamefont {Nardini}}, \bibinfo {author} {\bibfnamefont {A.}~\bibnamefont {Solon}}, \bibinfo {author} {\bibfnamefont {J.}~\bibnamefont {Tailleur}},\ and\ \bibinfo {author} {\bibfnamefont {F.}~\bibnamefont {Van~Wijland}},\ }\bibfield  {title} {\bibinfo {title} {Fluctuation-induced phase separation in metric and topological models of collective motion},\ }\href {https://doi.org/https://doi.org/10.1103/PhysRevLett.126.148001} {\bibfield  {journal} {\bibinfo  {journal} {Physical Review Letters}\ }\textbf {\bibinfo {volume} {126}},\ \bibinfo {pages} {148001} (\bibinfo {year} {2021})}\BibitemShut {NoStop}%
\bibitem [{\citenamefont {Martin}\ \emph {et~al.}(2024)\citenamefont {Martin}, \citenamefont {Spera}, \citenamefont {Chaté}, \citenamefont {Duclut}, \citenamefont {Nardini}, \citenamefont {Tailleur},\ and\ \citenamefont {van Wijland}}]{Martin_2024}%
  \BibitemOpen
  \bibfield  {author} {\bibinfo {author} {\bibfnamefont {D.}~\bibnamefont {Martin}}, \bibinfo {author} {\bibfnamefont {G.}~\bibnamefont {Spera}}, \bibinfo {author} {\bibfnamefont {H.}~\bibnamefont {Chaté}}, \bibinfo {author} {\bibfnamefont {C.}~\bibnamefont {Duclut}}, \bibinfo {author} {\bibfnamefont {C.}~\bibnamefont {Nardini}}, \bibinfo {author} {\bibfnamefont {J.}~\bibnamefont {Tailleur}},\ and\ \bibinfo {author} {\bibfnamefont {F.}~\bibnamefont {van Wijland}},\ }\bibfield  {title} {\bibinfo {title} {Fluctuation-induced first order transition to collective motion},\ }\href {https://doi.org/10.1088/1742-5468/ad6428} {\bibfield  {journal} {\bibinfo  {journal} {Journal of Statistical Mechanics: Theory and Experiment}\ }\textbf {\bibinfo {volume} {2024}},\ \bibinfo {pages} {084003} (\bibinfo {year} {2024})}\BibitemShut {NoStop}%
\bibitem [{\citenamefont {Rahmani}\ \emph {et~al.}(2021)\citenamefont {Rahmani}, \citenamefont {Peruani},\ and\ \citenamefont {Romanczuk}}]{rahmani2021topological}%
  \BibitemOpen
  \bibfield  {author} {\bibinfo {author} {\bibfnamefont {P.}~\bibnamefont {Rahmani}}, \bibinfo {author} {\bibfnamefont {F.}~\bibnamefont {Peruani}},\ and\ \bibinfo {author} {\bibfnamefont {P.}~\bibnamefont {Romanczuk}},\ }\bibfield  {title} {\bibinfo {title} {Topological flocking models in spatially heterogeneous environments},\ }\href {https://doi.org/https://doi.org/10.1038/s42005-021-00708-y} {\bibfield  {journal} {\bibinfo  {journal} {Communications Physics}\ }\textbf {\bibinfo {volume} {4}},\ \bibinfo {pages} {206} (\bibinfo {year} {2021})}\BibitemShut {NoStop}%
\bibitem [{\citenamefont {Shi}\ \emph {et~al.}(2022)\citenamefont {Shi}, \citenamefont {Du}, \citenamefont {Huang},\ and\ \citenamefont {Guo}}]{shi2022collective}%
  \BibitemOpen
  \bibfield  {author} {\bibinfo {author} {\bibfnamefont {H.-D.}\ \bibnamefont {Shi}}, \bibinfo {author} {\bibfnamefont {L.-C.}\ \bibnamefont {Du}}, \bibinfo {author} {\bibfnamefont {F.-J.}\ \bibnamefont {Huang}},\ and\ \bibinfo {author} {\bibfnamefont {W.}~\bibnamefont {Guo}},\ }\bibfield  {title} {\bibinfo {title} {Collective topological active particles: Non-ergodic superdiffusion and ageing in complex environments},\ }\href {https://doi.org/https://doi.org/10.1016/j.chaos.2022.111935} {\bibfield  {journal} {\bibinfo  {journal} {Chaos, Solitons \& Fractals}\ }\textbf {\bibinfo {volume} {157}},\ \bibinfo {pages} {111935} (\bibinfo {year} {2022})}\BibitemShut {NoStop}%
\bibitem [{\citenamefont {Ito}\ and\ \citenamefont {Uchida}(2024)}]{ito2024boltzmann}%
  \BibitemOpen
  \bibfield  {author} {\bibinfo {author} {\bibfnamefont {S.}~\bibnamefont {Ito}}\ and\ \bibinfo {author} {\bibfnamefont {N.}~\bibnamefont {Uchida}},\ }\bibfield  {title} {\bibinfo {title} {Boltzmann approach to collective motion via non-local visual interaction},\ }\href {https://doi.org/10.48550/arXiv.2408.09917} {\bibfield  {journal} {\bibinfo  {journal} {arXiv preprint arXiv:2408.09917}\ } (\bibinfo {year} {2024})}\BibitemShut {NoStop}%
\bibitem [{\citenamefont {Solon}\ \emph {et~al.}(2015{\natexlab{a}})\citenamefont {Solon}, \citenamefont {Chat{\'e}},\ and\ \citenamefont {Tailleur}}]{solon2015phase}%
  \BibitemOpen
  \bibfield  {author} {\bibinfo {author} {\bibfnamefont {A.~P.}\ \bibnamefont {Solon}}, \bibinfo {author} {\bibfnamefont {H.}~\bibnamefont {Chat{\'e}}},\ and\ \bibinfo {author} {\bibfnamefont {J.}~\bibnamefont {Tailleur}},\ }\bibfield  {title} {\bibinfo {title} {From phase to microphase separation in flocking models: The essential role of nonequilibrium fluctuations},\ }\href {https://doi.org/https://doi.org/10.1103/PhysRevLett.114.068101} {\bibfield  {journal} {\bibinfo  {journal} {Physical review letters}\ }\textbf {\bibinfo {volume} {114}},\ \bibinfo {pages} {068101} (\bibinfo {year} {2015}{\natexlab{a}})}\BibitemShut {NoStop}%
\bibitem [{\citenamefont {Duan}\ \emph {et~al.}(2021)\citenamefont {Duan}, \citenamefont {Mahault}, \citenamefont {Ma}, \citenamefont {Shi},\ and\ \citenamefont {Chat{\'e}}}]{duan2021breakdown}%
  \BibitemOpen
  \bibfield  {author} {\bibinfo {author} {\bibfnamefont {Y.}~\bibnamefont {Duan}}, \bibinfo {author} {\bibfnamefont {B.}~\bibnamefont {Mahault}}, \bibinfo {author} {\bibfnamefont {Y.-q.}\ \bibnamefont {Ma}}, \bibinfo {author} {\bibfnamefont {X.-q.}\ \bibnamefont {Shi}},\ and\ \bibinfo {author} {\bibfnamefont {H.}~\bibnamefont {Chat{\'e}}},\ }\bibfield  {title} {\bibinfo {title} {Breakdown of ergodicity and self-averaging in polar flocks with quenched disorder},\ }\href {https://doi.org/https://doi.org/10.1103/PhysRevLett.126.178001} {\bibfield  {journal} {\bibinfo  {journal} {Physical Review Letters}\ }\textbf {\bibinfo {volume} {126}},\ \bibinfo {pages} {178001} (\bibinfo {year} {2021})}\BibitemShut {NoStop}%
\bibitem [{\citenamefont {Bertrand}\ and\ \citenamefont {Lee}(2022)}]{bertrand2022diversity}%
  \BibitemOpen
  \bibfield  {author} {\bibinfo {author} {\bibfnamefont {T.}~\bibnamefont {Bertrand}}\ and\ \bibinfo {author} {\bibfnamefont {C.~F.}\ \bibnamefont {Lee}},\ }\bibfield  {title} {\bibinfo {title} {Diversity of phase transitions and phase separations in active fluids},\ }\href {https://doi.org/https://doi.org/10.1103/PhysRevResearch.4.L022046} {\bibfield  {journal} {\bibinfo  {journal} {Physical Review Research}\ }\textbf {\bibinfo {volume} {4}},\ \bibinfo {pages} {L022046} (\bibinfo {year} {2022})}\BibitemShut {NoStop}%
\bibitem [{\citenamefont {Jentsch}\ and\ \citenamefont {Lee}(2023)}]{jentsch2023critical}%
  \BibitemOpen
  \bibfield  {author} {\bibinfo {author} {\bibfnamefont {P.}~\bibnamefont {Jentsch}}\ and\ \bibinfo {author} {\bibfnamefont {C.~F.}\ \bibnamefont {Lee}},\ }\bibfield  {title} {\bibinfo {title} {Critical phenomena in compressible polar active fluids: Dynamical and functional renormalization group studies},\ }\href {https://doi.org/https://doi.org/10.1103/PhysRevResearch.5.023061} {\bibfield  {journal} {\bibinfo  {journal} {Physical Review Research}\ }\textbf {\bibinfo {volume} {5}},\ \bibinfo {pages} {023061} (\bibinfo {year} {2023})}\BibitemShut {NoStop}%
\bibitem [{\citenamefont {Agranov}\ \emph {et~al.}(2024)\citenamefont {Agranov}, \citenamefont {Jack}, \citenamefont {Cates},\ and\ \citenamefont {Fodor}}]{agranov2024thermodynamically}%
  \BibitemOpen
  \bibfield  {author} {\bibinfo {author} {\bibfnamefont {T.}~\bibnamefont {Agranov}}, \bibinfo {author} {\bibfnamefont {R.~L.}\ \bibnamefont {Jack}}, \bibinfo {author} {\bibfnamefont {M.~E.}\ \bibnamefont {Cates}},\ and\ \bibinfo {author} {\bibfnamefont {{\'E}.}~\bibnamefont {Fodor}},\ }\bibfield  {title} {\bibinfo {title} {Thermodynamically consistent flocking: from discontinuous to continuous transitions},\ }\href {https://doi.org/10.1088/1367-2630/ad4dd6} {\bibfield  {journal} {\bibinfo  {journal} {New Journal of Physics}\ }\textbf {\bibinfo {volume} {26}},\ \bibinfo {pages} {063006} (\bibinfo {year} {2024})}\BibitemShut {NoStop}%
\bibitem [{\citenamefont {Gautrais}\ \emph {et~al.}(2012)\citenamefont {Gautrais}, \citenamefont {Ginelli}, \citenamefont {Fournier}, \citenamefont {Blanco}, \citenamefont {Soria}, \citenamefont {Chat{\'e}},\ and\ \citenamefont {Theraulaz}}]{Gautrais2012DecipheringII}%
  \BibitemOpen
  \bibfield  {author} {\bibinfo {author} {\bibfnamefont {J.}~\bibnamefont {Gautrais}}, \bibinfo {author} {\bibfnamefont {F.}~\bibnamefont {Ginelli}}, \bibinfo {author} {\bibfnamefont {R.}~\bibnamefont {Fournier}}, \bibinfo {author} {\bibfnamefont {S.}~\bibnamefont {Blanco}}, \bibinfo {author} {\bibfnamefont {M.}~\bibnamefont {Soria}}, \bibinfo {author} {\bibfnamefont {H.}~\bibnamefont {Chat{\'e}}},\ and\ \bibinfo {author} {\bibfnamefont {G.}~\bibnamefont {Theraulaz}},\ }\bibfield  {title} {\bibinfo {title} {Deciphering interactions in moving animal groups},\ }\href {https://api.semanticscholar.org/CorpusID:5711875} {\bibfield  {journal} {\bibinfo  {journal} {PLoS Computational Biology}\ }\textbf {\bibinfo {volume} {8}} (\bibinfo {year} {2012})}\BibitemShut {NoStop}%
\bibitem [{\citenamefont {Camperi}\ \emph {et~al.}(2012)\citenamefont {Camperi}, \citenamefont {Cavagna}, \citenamefont {Giardina}, \citenamefont {Parisi},\ and\ \citenamefont {Silvestri}}]{camperi2012spatially}%
  \BibitemOpen
  \bibfield  {author} {\bibinfo {author} {\bibfnamefont {M.}~\bibnamefont {Camperi}}, \bibinfo {author} {\bibfnamefont {A.}~\bibnamefont {Cavagna}}, \bibinfo {author} {\bibfnamefont {I.}~\bibnamefont {Giardina}}, \bibinfo {author} {\bibfnamefont {G.}~\bibnamefont {Parisi}},\ and\ \bibinfo {author} {\bibfnamefont {E.}~\bibnamefont {Silvestri}},\ }\bibfield  {title} {\bibinfo {title} {Spatially balanced topological interaction grants optimal cohesion in flocking models},\ }\href {https://doi.org/https://doi.org/10.1098/rsfs.2012.0026} {\bibfield  {journal} {\bibinfo  {journal} {Interface focus}\ }\textbf {\bibinfo {volume} {2}},\ \bibinfo {pages} {715} (\bibinfo {year} {2012})}\BibitemShut {NoStop}%
\bibitem [{\citenamefont {Ginelli}\ \emph {et~al.}(2015)\citenamefont {Ginelli}, \citenamefont {Peruani}, \citenamefont {Pillot}, \citenamefont {Chat{\'e}}, \citenamefont {Theraulaz},\ and\ \citenamefont {Bon}}]{ginelli2015intermittent}%
  \BibitemOpen
  \bibfield  {author} {\bibinfo {author} {\bibfnamefont {F.}~\bibnamefont {Ginelli}}, \bibinfo {author} {\bibfnamefont {F.}~\bibnamefont {Peruani}}, \bibinfo {author} {\bibfnamefont {M.-H.}\ \bibnamefont {Pillot}}, \bibinfo {author} {\bibfnamefont {H.}~\bibnamefont {Chat{\'e}}}, \bibinfo {author} {\bibfnamefont {G.}~\bibnamefont {Theraulaz}},\ and\ \bibinfo {author} {\bibfnamefont {R.}~\bibnamefont {Bon}},\ }\bibfield  {title} {\bibinfo {title} {Intermittent collective dynamics emerge from conflicting imperatives in sheep herds},\ }\href {https://doi.org/https://doi.org/10.1073/pnas.1503749112} {\bibfield  {journal} {\bibinfo  {journal} {Proceedings of the National Academy of Sciences}\ }\textbf {\bibinfo {volume} {112}},\ \bibinfo {pages} {12729} (\bibinfo {year} {2015})}\BibitemShut {NoStop}%
\bibitem [{\citenamefont {Pearce}\ \emph {et~al.}(2014)\citenamefont {Pearce}, \citenamefont {Miller}, \citenamefont {Rowlands},\ and\ \citenamefont {Turner}}]{pearce2014role}%
  \BibitemOpen
  \bibfield  {author} {\bibinfo {author} {\bibfnamefont {D.~J.}\ \bibnamefont {Pearce}}, \bibinfo {author} {\bibfnamefont {A.~M.}\ \bibnamefont {Miller}}, \bibinfo {author} {\bibfnamefont {G.}~\bibnamefont {Rowlands}},\ and\ \bibinfo {author} {\bibfnamefont {M.~S.}\ \bibnamefont {Turner}},\ }\bibfield  {title} {\bibinfo {title} {Role of projection in the control of bird flocks},\ }\href {https://doi.org/https://doi.org/10.1073/pnas.1402202111} {\bibfield  {journal} {\bibinfo  {journal} {Proceedings of the National Academy of Sciences}\ }\textbf {\bibinfo {volume} {111}},\ \bibinfo {pages} {10422} (\bibinfo {year} {2014})}\BibitemShut {NoStop}%
\bibitem [{\citenamefont {Jiang}\ \emph {et~al.}(2017)\citenamefont {Jiang}, \citenamefont {Giuggioli}, \citenamefont {Perna}, \citenamefont {Escobedo}, \citenamefont {Lecheval}, \citenamefont {Sire}, \citenamefont {Han},\ and\ \citenamefont {Theraulaz}}]{Jiang2017IdentifyingIN}%
  \BibitemOpen
  \bibfield  {author} {\bibinfo {author} {\bibfnamefont {L.}~\bibnamefont {Jiang}}, \bibinfo {author} {\bibfnamefont {L.}~\bibnamefont {Giuggioli}}, \bibinfo {author} {\bibfnamefont {A.}~\bibnamefont {Perna}}, \bibinfo {author} {\bibfnamefont {R.}~\bibnamefont {Escobedo}}, \bibinfo {author} {\bibfnamefont {V.}~\bibnamefont {Lecheval}}, \bibinfo {author} {\bibfnamefont {C.}~\bibnamefont {Sire}}, \bibinfo {author} {\bibfnamefont {Z.}~\bibnamefont {Han}},\ and\ \bibinfo {author} {\bibfnamefont {G.}~\bibnamefont {Theraulaz}},\ }\bibfield  {title} {\bibinfo {title} {Identifying influential neighbors in animal flocking},\ }\href {https://api.semanticscholar.org/CorpusID:3798208} {\bibfield  {journal} {\bibinfo  {journal} {PLoS Computational Biology}\ }\textbf {\bibinfo {volume} {13}} (\bibinfo {year} {2017})}\BibitemShut {NoStop}%
\bibitem [{\citenamefont {Jhawar}\ \emph {et~al.}(2020)\citenamefont {Jhawar}, \citenamefont {Morris}, \citenamefont {Amith-Kumar}, \citenamefont {Danny~Raj}, \citenamefont {Rogers}, \citenamefont {Rajendran},\ and\ \citenamefont {Guttal}}]{jhawar2020noise}%
  \BibitemOpen
  \bibfield  {author} {\bibinfo {author} {\bibfnamefont {J.}~\bibnamefont {Jhawar}}, \bibinfo {author} {\bibfnamefont {R.~G.}\ \bibnamefont {Morris}}, \bibinfo {author} {\bibfnamefont {U.}~\bibnamefont {Amith-Kumar}}, \bibinfo {author} {\bibfnamefont {M.}~\bibnamefont {Danny~Raj}}, \bibinfo {author} {\bibfnamefont {T.}~\bibnamefont {Rogers}}, \bibinfo {author} {\bibfnamefont {H.}~\bibnamefont {Rajendran}},\ and\ \bibinfo {author} {\bibfnamefont {V.}~\bibnamefont {Guttal}},\ }\bibfield  {title} {\bibinfo {title} {Noise-induced schooling of fish},\ }\href {https://doi.org/https://doi.org/10.1038/s41567-020-0787-y} {\bibfield  {journal} {\bibinfo  {journal} {Nature Physics}\ }\textbf {\bibinfo {volume} {16}},\ \bibinfo {pages} {488} (\bibinfo {year} {2020})}\BibitemShut {NoStop}%
\bibitem [{\citenamefont {Moussa{\"\i}d}\ \emph {et~al.}(2011)\citenamefont {Moussa{\"\i}d}, \citenamefont {Helbing},\ and\ \citenamefont {Theraulaz}}]{moussaid2011simple}%
  \BibitemOpen
  \bibfield  {author} {\bibinfo {author} {\bibfnamefont {M.}~\bibnamefont {Moussa{\"\i}d}}, \bibinfo {author} {\bibfnamefont {D.}~\bibnamefont {Helbing}},\ and\ \bibinfo {author} {\bibfnamefont {G.}~\bibnamefont {Theraulaz}},\ }\bibfield  {title} {\bibinfo {title} {How simple rules determine pedestrian behavior and crowd disasters},\ }\href {https://doi.org/https://doi.org/10.1073/pnas.1016507108} {\bibfield  {journal} {\bibinfo  {journal} {Proceedings of the National Academy of Sciences}\ }\textbf {\bibinfo {volume} {108}},\ \bibinfo {pages} {6884} (\bibinfo {year} {2011})}\BibitemShut {NoStop}%
\bibitem [{\citenamefont {Peshkov}\ \emph {et~al.}(2012)\citenamefont {Peshkov}, \citenamefont {Ngo}, \citenamefont {Bertin}, \citenamefont {Chat{\'e}},\ and\ \citenamefont {Ginelli}}]{peshkov2012continuous}%
  \BibitemOpen
  \bibfield  {author} {\bibinfo {author} {\bibfnamefont {A.}~\bibnamefont {Peshkov}}, \bibinfo {author} {\bibfnamefont {S.}~\bibnamefont {Ngo}}, \bibinfo {author} {\bibfnamefont {E.}~\bibnamefont {Bertin}}, \bibinfo {author} {\bibfnamefont {H.}~\bibnamefont {Chat{\'e}}},\ and\ \bibinfo {author} {\bibfnamefont {F.}~\bibnamefont {Ginelli}},\ }\bibfield  {title} {\bibinfo {title} {Continuous theory of active matter systems with metric-free interactions},\ }\href {https://doi.org/https://doi.org/10.1103/PhysRevLett.109.098101} {\bibfield  {journal} {\bibinfo  {journal} {Physical review letters}\ }\textbf {\bibinfo {volume} {109}},\ \bibinfo {pages} {098101} (\bibinfo {year} {2012})}\BibitemShut {NoStop}%
\bibitem [{\citenamefont {Chou}\ \emph {et~al.}(2012)\citenamefont {Chou}, \citenamefont {Wolfe},\ and\ \citenamefont {Ihle}}]{chou2012kinetic}%
  \BibitemOpen
  \bibfield  {author} {\bibinfo {author} {\bibfnamefont {Y.-L.}\ \bibnamefont {Chou}}, \bibinfo {author} {\bibfnamefont {R.}~\bibnamefont {Wolfe}},\ and\ \bibinfo {author} {\bibfnamefont {T.}~\bibnamefont {Ihle}},\ }\bibfield  {title} {\bibinfo {title} {Kinetic theory for systems of self-propelled particles with metric-free interactions},\ }\href {https://doi.org/https://doi.org/10.1103/PhysRevE.86.021120} {\bibfield  {journal} {\bibinfo  {journal} {Physical Review E—Statistical, Nonlinear, and Soft Matter Physics}\ }\textbf {\bibinfo {volume} {86}},\ \bibinfo {pages} {021120} (\bibinfo {year} {2012})}\BibitemShut {NoStop}%
\bibitem [{\citenamefont {Packard}\ and\ \citenamefont {Sussman}(2024)}]{packard2024banded}%
  \BibitemOpen
  \bibfield  {author} {\bibinfo {author} {\bibfnamefont {C.~R.}\ \bibnamefont {Packard}}\ and\ \bibinfo {author} {\bibfnamefont {D.~M.}\ \bibnamefont {Sussman}},\ }\bibfield  {title} {\bibinfo {title} {Banded phases in topological flocks},\ }\href {https://doi.org/10.48550/arXiv.2409.05198} {\bibfield  {journal} {\bibinfo  {journal} {arXiv preprint arXiv:2409.05198}\ } (\bibinfo {year} {2024})}\BibitemShut {NoStop}%
\bibitem [{\citenamefont {Maity}\ and\ \citenamefont {Morin}(2023)}]{maity2023spontaneous}%
  \BibitemOpen
  \bibfield  {author} {\bibinfo {author} {\bibfnamefont {S.}~\bibnamefont {Maity}}\ and\ \bibinfo {author} {\bibfnamefont {A.}~\bibnamefont {Morin}},\ }\bibfield  {title} {\bibinfo {title} {Spontaneous demixing of binary colloidal flocks},\ }\href {https://doi.org/https://doi.org/10.1103/PhysRevLett.131.178304} {\bibfield  {journal} {\bibinfo  {journal} {Phys. Rev. Lett.}\ }\textbf {\bibinfo {volume} {131}},\ \bibinfo {pages} {178304} (\bibinfo {year} {2023})}\BibitemShut {NoStop}%
\bibitem [{\citenamefont {Soto}\ and\ \citenamefont {Golestanian}(2014)}]{soto2014self}%
  \BibitemOpen
  \bibfield  {author} {\bibinfo {author} {\bibfnamefont {R.}~\bibnamefont {Soto}}\ and\ \bibinfo {author} {\bibfnamefont {R.}~\bibnamefont {Golestanian}},\ }\bibfield  {title} {\bibinfo {title} {Self-assembly of catalytically active colloidal molecules: tailoring activity through surface chemistry},\ }\href {https://doi.org/https://doi.org/10.1103/PhysRevLett.112.068301} {\bibfield  {journal} {\bibinfo  {journal} {Physical review letters}\ }\textbf {\bibinfo {volume} {112}},\ \bibinfo {pages} {068301} (\bibinfo {year} {2014})}\BibitemShut {NoStop}%
\bibitem [{\citenamefont {Lavergne}\ \emph {et~al.}(2019)\citenamefont {Lavergne}, \citenamefont {Wendehenne}, \citenamefont {B{\"a}uerle},\ and\ \citenamefont {Bechinger}}]{lavergne2019Science}%
  \BibitemOpen
  \bibfield  {author} {\bibinfo {author} {\bibfnamefont {F.~A.}\ \bibnamefont {Lavergne}}, \bibinfo {author} {\bibfnamefont {H.}~\bibnamefont {Wendehenne}}, \bibinfo {author} {\bibfnamefont {T.}~\bibnamefont {B{\"a}uerle}},\ and\ \bibinfo {author} {\bibfnamefont {C.}~\bibnamefont {Bechinger}},\ }\bibfield  {title} {\bibinfo {title} {Group formation and cohesion of active particles with visual perception--dependent motility},\ }\href {https://doi.org/10.1126/science.aau5347} {\bibfield  {journal} {\bibinfo  {journal} {Science}\ }\textbf {\bibinfo {volume} {364}},\ \bibinfo {pages} {70} (\bibinfo {year} {2019})}\BibitemShut {NoStop}%
\bibitem [{\citenamefont {Fruchart}\ \emph {et~al.}(2021)\citenamefont {Fruchart}, \citenamefont {Hanai}, \citenamefont {Littlewood},\ and\ \citenamefont {Vitelli}}]{fruchart2021non}%
  \BibitemOpen
  \bibfield  {author} {\bibinfo {author} {\bibfnamefont {M.}~\bibnamefont {Fruchart}}, \bibinfo {author} {\bibfnamefont {R.}~\bibnamefont {Hanai}}, \bibinfo {author} {\bibfnamefont {P.~B.}\ \bibnamefont {Littlewood}},\ and\ \bibinfo {author} {\bibfnamefont {V.}~\bibnamefont {Vitelli}},\ }\bibfield  {title} {\bibinfo {title} {Non-reciprocal phase transitions},\ }\href {https://doi.org/https://doi.org/10.1038/s41586-021-03375-9} {\bibfield  {journal} {\bibinfo  {journal} {Nature}\ }\textbf {\bibinfo {volume} {592}},\ \bibinfo {pages} {363} (\bibinfo {year} {2021})}\BibitemShut {NoStop}%
\bibitem [{\citenamefont {Saha}\ \emph {et~al.}(2020)\citenamefont {Saha}, \citenamefont {Agudo-Canalejo},\ and\ \citenamefont {Golestanian}}]{saha2020scalar}%
  \BibitemOpen
  \bibfield  {author} {\bibinfo {author} {\bibfnamefont {S.}~\bibnamefont {Saha}}, \bibinfo {author} {\bibfnamefont {J.}~\bibnamefont {Agudo-Canalejo}},\ and\ \bibinfo {author} {\bibfnamefont {R.}~\bibnamefont {Golestanian}},\ }\bibfield  {title} {\bibinfo {title} {{Scalar active mixtures: The nonreciprocal Cahn-Hilliard model}},\ }\href {https://doi.org/10.1103/PhysRevX.10.041009} {\bibfield  {journal} {\bibinfo  {journal} {Phys. Rev. X}\ }\textbf {\bibinfo {volume} {10}},\ \bibinfo {pages} {041009} (\bibinfo {year} {2020})}\BibitemShut {NoStop}%
\bibitem [{\citenamefont {You}\ \emph {et~al.}(2020)\citenamefont {You}, \citenamefont {Baskaran},\ and\ \citenamefont {Marchetti}}]{you2020nonreciprocity}%
  \BibitemOpen
  \bibfield  {author} {\bibinfo {author} {\bibfnamefont {Z.}~\bibnamefont {You}}, \bibinfo {author} {\bibfnamefont {A.}~\bibnamefont {Baskaran}},\ and\ \bibinfo {author} {\bibfnamefont {M.~C.}\ \bibnamefont {Marchetti}},\ }\bibfield  {title} {\bibinfo {title} {Nonreciprocity as a generic route to traveling states},\ }\href {https://doi.org/10.1073/pnas.2010318117} {\bibfield  {journal} {\bibinfo  {journal} {Proc. Natl. Acad. Sci. USA.}\ }\textbf {\bibinfo {volume} {117}},\ \bibinfo {pages} {19767} (\bibinfo {year} {2020})}\BibitemShut {NoStop}%
\bibitem [{\citenamefont {Duan}\ \emph {et~al.}(2023)\citenamefont {Duan}, \citenamefont {Agudo-Canalejo}, \citenamefont {Golestanian},\ and\ \citenamefont {Mahault}}]{duan2023dynamical}%
  \BibitemOpen
  \bibfield  {author} {\bibinfo {author} {\bibfnamefont {Y.}~\bibnamefont {Duan}}, \bibinfo {author} {\bibfnamefont {J.}~\bibnamefont {Agudo-Canalejo}}, \bibinfo {author} {\bibfnamefont {R.}~\bibnamefont {Golestanian}},\ and\ \bibinfo {author} {\bibfnamefont {B.}~\bibnamefont {Mahault}},\ }\bibfield  {title} {\bibinfo {title} {{Dynamical Pattern Formation without Self-Attraction in Quorum-Sensing Active Matter: The Interplay between Nonreciprocity and Motility}},\ }\href {https://doi.org/https://doi.org/10.1103/PhysRevLett.131.148301} {\bibfield  {journal} {\bibinfo  {journal} {Phys. Rev. Lett.}\ }\textbf {\bibinfo {volume} {131}},\ \bibinfo {pages} {148301} (\bibinfo {year} {2023})}\BibitemShut {NoStop}%
\bibitem [{\citenamefont {Dinelli}\ \emph {et~al.}(2023)\citenamefont {Dinelli}, \citenamefont {O'Byrne}, \citenamefont {Curatolo}, \citenamefont {Zhao}, \citenamefont {Sollich},\ and\ \citenamefont {Tailleur}}]{dinelli2023non}%
  \BibitemOpen
  \bibfield  {author} {\bibinfo {author} {\bibfnamefont {A.}~\bibnamefont {Dinelli}}, \bibinfo {author} {\bibfnamefont {J.}~\bibnamefont {O'Byrne}}, \bibinfo {author} {\bibfnamefont {A.}~\bibnamefont {Curatolo}}, \bibinfo {author} {\bibfnamefont {Y.}~\bibnamefont {Zhao}}, \bibinfo {author} {\bibfnamefont {P.}~\bibnamefont {Sollich}},\ and\ \bibinfo {author} {\bibfnamefont {J.}~\bibnamefont {Tailleur}},\ }\bibfield  {title} {\bibinfo {title} {Non-reciprocity across scales in active mixtures},\ }\href {https://doi.org/10.1038/s41467-023-42713-5} {\bibfield  {journal} {\bibinfo  {journal} {Nat. Commun.}\ }\textbf {\bibinfo {volume} {14}},\ \bibinfo {pages} {7035} (\bibinfo {year} {2023})}\BibitemShut {NoStop}%
\bibitem [{\citenamefont {Chen}\ \emph {et~al.}(2024)\citenamefont {Chen}, \citenamefont {Lei}, \citenamefont {Xiang}, \citenamefont {Duan}, \citenamefont {Peng},\ and\ \citenamefont {Zhang}}]{chen2024emergent}%
  \BibitemOpen
  \bibfield  {author} {\bibinfo {author} {\bibfnamefont {J.}~\bibnamefont {Chen}}, \bibinfo {author} {\bibfnamefont {X.}~\bibnamefont {Lei}}, \bibinfo {author} {\bibfnamefont {Y.}~\bibnamefont {Xiang}}, \bibinfo {author} {\bibfnamefont {M.}~\bibnamefont {Duan}}, \bibinfo {author} {\bibfnamefont {X.}~\bibnamefont {Peng}},\ and\ \bibinfo {author} {\bibfnamefont {H.}~\bibnamefont {Zhang}},\ }\bibfield  {title} {\bibinfo {title} {Emergent chirality and hyperuniformity in an active mixture with nonreciprocal interactions},\ }\href {https://doi.org/https://doi.org/10.1103/PhysRevLett.132.118301} {\bibfield  {journal} {\bibinfo  {journal} {Physical Review Letters}\ }\textbf {\bibinfo {volume} {132}},\ \bibinfo {pages} {118301} (\bibinfo {year} {2024})}\BibitemShut {NoStop}%
\bibitem [{\citenamefont {Kant}\ \emph {et~al.}(2024)\citenamefont {Kant}, \citenamefont {Gupta}, \citenamefont {Soni}, \citenamefont {Sood},\ and\ \citenamefont {Ramaswamy}}]{kant2024bulk}%
  \BibitemOpen
  \bibfield  {author} {\bibinfo {author} {\bibfnamefont {R.}~\bibnamefont {Kant}}, \bibinfo {author} {\bibfnamefont {R.~K.}\ \bibnamefont {Gupta}}, \bibinfo {author} {\bibfnamefont {H.}~\bibnamefont {Soni}}, \bibinfo {author} {\bibfnamefont {A.}~\bibnamefont {Sood}},\ and\ \bibinfo {author} {\bibfnamefont {S.}~\bibnamefont {Ramaswamy}},\ }\bibfield  {title} {\bibinfo {title} {Bulk condensation by an active interface},\ }\href {https://doi.org/https://doi.org/10.1103/PhysRevLett.133.208301} {\bibfield  {journal} {\bibinfo  {journal} {Physical Review Letters}\ }\textbf {\bibinfo {volume} {133}},\ \bibinfo {pages} {208301} (\bibinfo {year} {2024})}\BibitemShut {NoStop}%
\bibitem [{\citenamefont {Chen}\ \emph {et~al.}(2017)\citenamefont {Chen}, \citenamefont {Patelli}, \citenamefont {Chat{\'e}}, \citenamefont {Ma},\ and\ \citenamefont {Shi}}]{chen2017fore}%
  \BibitemOpen
  \bibfield  {author} {\bibinfo {author} {\bibfnamefont {Q.-s.}\ \bibnamefont {Chen}}, \bibinfo {author} {\bibfnamefont {A.}~\bibnamefont {Patelli}}, \bibinfo {author} {\bibfnamefont {H.}~\bibnamefont {Chat{\'e}}}, \bibinfo {author} {\bibfnamefont {Y.-q.}\ \bibnamefont {Ma}},\ and\ \bibinfo {author} {\bibfnamefont {X.-q.}\ \bibnamefont {Shi}},\ }\bibfield  {title} {\bibinfo {title} {Fore-aft asymmetric flocking},\ }\href {https://doi.org/https://doi.org/10.1103/PhysRevE.96.020601} {\bibfield  {journal} {\bibinfo  {journal} {Physical Review E}\ }\textbf {\bibinfo {volume} {96}},\ \bibinfo {pages} {020601} (\bibinfo {year} {2017})}\BibitemShut {NoStop}%
\bibitem [{\citenamefont {Martin}\ \emph {et~al.}(2023)\citenamefont {Martin}, \citenamefont {Seara}, \citenamefont {Avni}, \citenamefont {Fruchart},\ and\ \citenamefont {Vitelli}}]{martin2023exact}%
  \BibitemOpen
  \bibfield  {author} {\bibinfo {author} {\bibfnamefont {D.}~\bibnamefont {Martin}}, \bibinfo {author} {\bibfnamefont {D.}~\bibnamefont {Seara}}, \bibinfo {author} {\bibfnamefont {Y.}~\bibnamefont {Avni}}, \bibinfo {author} {\bibfnamefont {M.}~\bibnamefont {Fruchart}},\ and\ \bibinfo {author} {\bibfnamefont {V.}~\bibnamefont {Vitelli}},\ }\bibfield  {title} {\bibinfo {title} {An exact model for the transition to collective motion in nonreciprocal active matter},\ }\href {https://doi.org/10.48550/arXiv.2307.08251} {\bibfield  {journal} {\bibinfo  {journal} {arXiv preprint arXiv:2307.08251}\ } (\bibinfo {year} {2023})}\BibitemShut {NoStop}%
\bibitem [{\citenamefont {Kreienkamp}\ and\ \citenamefont {Klapp}(2024)}]{kreienkamp2024non}%
  \BibitemOpen
  \bibfield  {author} {\bibinfo {author} {\bibfnamefont {K.~L.}\ \bibnamefont {Kreienkamp}}\ and\ \bibinfo {author} {\bibfnamefont {S.~H.}\ \bibnamefont {Klapp}},\ }\bibfield  {title} {\bibinfo {title} {Non-reciprocal alignment induces asymmetric clustering in active repulsive mixtures},\ }\href {https://doi.org/10.48550/arXiv.2403.19291} {\bibfield  {journal} {\bibinfo  {journal} {arXiv preprint arXiv:2403.19291}\ } (\bibinfo {year} {2024})}\BibitemShut {NoStop}%
\bibitem [{\citenamefont {Mangeat}\ \emph {et~al.}(2024)\citenamefont {Mangeat}, \citenamefont {Chatterjee}, \citenamefont {Noh},\ and\ \citenamefont {Rieger}}]{mangeat2024emergent}%
  \BibitemOpen
  \bibfield  {author} {\bibinfo {author} {\bibfnamefont {M.}~\bibnamefont {Mangeat}}, \bibinfo {author} {\bibfnamefont {S.}~\bibnamefont {Chatterjee}}, \bibinfo {author} {\bibfnamefont {J.~D.}\ \bibnamefont {Noh}},\ and\ \bibinfo {author} {\bibfnamefont {H.}~\bibnamefont {Rieger}},\ }\bibfield  {title} {\bibinfo {title} {Emergent complex phases in a discrete flocking model with reciprocal and non-reciprocal interactions},\ }\href {https://doi.org/10.48550/arXiv.2412.02501} {\bibfield  {journal} {\bibinfo  {journal} {arXiv preprint arXiv:2412.02501}\ } (\bibinfo {year} {2024})}\BibitemShut {NoStop}%
\bibitem [{\citenamefont {Bera}\ and\ \citenamefont {Sood}(2020)}]{bera2020motile}%
  \BibitemOpen
  \bibfield  {author} {\bibinfo {author} {\bibfnamefont {P.~K.}\ \bibnamefont {Bera}}\ and\ \bibinfo {author} {\bibfnamefont {A.}~\bibnamefont {Sood}},\ }\bibfield  {title} {\bibinfo {title} {Motile dissenters disrupt the flocking of active granular matter},\ }\href {https://doi.org/https://doi.org/10.1103/PhysRevE.101.052615} {\bibfield  {journal} {\bibinfo  {journal} {Physical Review E}\ }\textbf {\bibinfo {volume} {101}},\ \bibinfo {pages} {052615} (\bibinfo {year} {2020})}\BibitemShut {NoStop}%
\bibitem [{\citenamefont {Yllanes}\ \emph {et~al.}(2017)\citenamefont {Yllanes}, \citenamefont {Leoni},\ and\ \citenamefont {Marchetti}}]{yllanes2017many}%
  \BibitemOpen
  \bibfield  {author} {\bibinfo {author} {\bibfnamefont {D.}~\bibnamefont {Yllanes}}, \bibinfo {author} {\bibfnamefont {M.}~\bibnamefont {Leoni}},\ and\ \bibinfo {author} {\bibfnamefont {M.}~\bibnamefont {Marchetti}},\ }\bibfield  {title} {\bibinfo {title} {How many dissenters does it take to disorder a flock?},\ }\href {https://doi.org/10.1088/1367-2630/aa8ed7} {\bibfield  {journal} {\bibinfo  {journal} {New Journal of Physics}\ }\textbf {\bibinfo {volume} {19}},\ \bibinfo {pages} {103026} (\bibinfo {year} {2017})}\BibitemShut {NoStop}%
\bibitem [{\citenamefont {Bertin}\ \emph {et~al.}(2009)\citenamefont {Bertin}, \citenamefont {Droz},\ and\ \citenamefont {Gr{\'e}goire}}]{bertin2009hydrodynamic}%
  \BibitemOpen
  \bibfield  {author} {\bibinfo {author} {\bibfnamefont {E.}~\bibnamefont {Bertin}}, \bibinfo {author} {\bibfnamefont {M.}~\bibnamefont {Droz}},\ and\ \bibinfo {author} {\bibfnamefont {G.}~\bibnamefont {Gr{\'e}goire}},\ }\bibfield  {title} {\bibinfo {title} {Hydrodynamic equations for self-propelled particles: microscopic derivation and stability analysis},\ }\href {https://doi.org/10.1088/1751-8113/42/44/445001} {\bibfield  {journal} {\bibinfo  {journal} {Journal of Physics A: Mathematical and Theoretical}\ }\textbf {\bibinfo {volume} {42}},\ \bibinfo {pages} {445001} (\bibinfo {year} {2009})}\BibitemShut {NoStop}%
\bibitem [{\citenamefont {Mishra}\ \emph {et~al.}(2010)\citenamefont {Mishra}, \citenamefont {Baskaran},\ and\ \citenamefont {Marchetti}}]{mishra2010fluctuations}%
  \BibitemOpen
  \bibfield  {author} {\bibinfo {author} {\bibfnamefont {S.}~\bibnamefont {Mishra}}, \bibinfo {author} {\bibfnamefont {A.}~\bibnamefont {Baskaran}},\ and\ \bibinfo {author} {\bibfnamefont {M.~C.}\ \bibnamefont {Marchetti}},\ }\bibfield  {title} {\bibinfo {title} {Fluctuations and pattern formation in self-propelled particles},\ }\href {https://doi.org/https://doi.org/10.1103/PhysRevE.81.061916} {\bibfield  {journal} {\bibinfo  {journal} {Physical Review E—Statistical, Nonlinear, and Soft Matter Physics}\ }\textbf {\bibinfo {volume} {81}},\ \bibinfo {pages} {061916} (\bibinfo {year} {2010})}\BibitemShut {NoStop}%
\bibitem [{\citenamefont {Miller}\ and\ \citenamefont {Toner}(2024)}]{miller2024following}%
  \BibitemOpen
  \bibfield  {author} {\bibinfo {author} {\bibfnamefont {M.}~\bibnamefont {Miller}}\ and\ \bibinfo {author} {\bibfnamefont {J.}~\bibnamefont {Toner}},\ }\bibfield  {title} {\bibinfo {title} {Following your nose: Autochemotaxis and other mechanisms for spinodal decomposition in flocks},\ }\href {https://doi.org/https://doi.org/10.1103/PhysRevLett.132.128301} {\bibfield  {journal} {\bibinfo  {journal} {Physical Review Letters}\ }\textbf {\bibinfo {volume} {132}},\ \bibinfo {pages} {128301} (\bibinfo {year} {2024})}\BibitemShut {NoStop}%
\bibitem [{\citenamefont {Solon}\ \emph {et~al.}(2015{\natexlab{b}})\citenamefont {Solon}, \citenamefont {Caussin}, \citenamefont {Bartolo}, \citenamefont {Chat{\'e}},\ and\ \citenamefont {Tailleur}}]{solon2015pattern}%
  \BibitemOpen
  \bibfield  {author} {\bibinfo {author} {\bibfnamefont {A.~P.}\ \bibnamefont {Solon}}, \bibinfo {author} {\bibfnamefont {J.-B.}\ \bibnamefont {Caussin}}, \bibinfo {author} {\bibfnamefont {D.}~\bibnamefont {Bartolo}}, \bibinfo {author} {\bibfnamefont {H.}~\bibnamefont {Chat{\'e}}},\ and\ \bibinfo {author} {\bibfnamefont {J.}~\bibnamefont {Tailleur}},\ }\bibfield  {title} {\bibinfo {title} {Pattern formation in flocking models: A hydrodynamic description},\ }\href {https://doi.org/https://doi.org/10.1103/PhysRevE.92.062111} {\bibfield  {journal} {\bibinfo  {journal} {Physical Review E}\ }\textbf {\bibinfo {volume} {92}},\ \bibinfo {pages} {062111} (\bibinfo {year} {2015}{\natexlab{b}})}\BibitemShut {NoStop}%
\bibitem [{\citenamefont {Mahault}(2018)}]{mahault2018outstanding}%
  \BibitemOpen
  \bibfield  {author} {\bibinfo {author} {\bibfnamefont {B.}~\bibnamefont {Mahault}},\ }\emph {\bibinfo {title} {Outstanding problems in the statistical physics of active matter}},\ \href {https://hal.science/tel-01880315/} {Ph.D. thesis},\ \bibinfo  {school} {Universit{\'e} Paris Saclay (COmUE)} (\bibinfo {year} {2018})\BibitemShut {NoStop}%
\bibitem [{\citenamefont {Peshkov}\ \emph {et~al.}(2014)\citenamefont {Peshkov}, \citenamefont {Bertin}, \citenamefont {Ginelli},\ and\ \citenamefont {Chat{\'e}}}]{peshkov2014boltzmann}%
  \BibitemOpen
  \bibfield  {author} {\bibinfo {author} {\bibfnamefont {A.}~\bibnamefont {Peshkov}}, \bibinfo {author} {\bibfnamefont {E.}~\bibnamefont {Bertin}}, \bibinfo {author} {\bibfnamefont {F.}~\bibnamefont {Ginelli}},\ and\ \bibinfo {author} {\bibfnamefont {H.}~\bibnamefont {Chat{\'e}}},\ }\bibfield  {title} {\bibinfo {title} {Boltzmann-ginzburg-landau approach for continuous descriptions of generic vicsek-like models},\ }\href {https://doi.org/https://doi.org/10.1140/epjst/e2014-02193-y} {\bibfield  {journal} {\bibinfo  {journal} {The European Physical Journal Special Topics}\ }\textbf {\bibinfo {volume} {223}},\ \bibinfo {pages} {1315} (\bibinfo {year} {2014})}\BibitemShut {NoStop}%
\bibitem [{\citenamefont {Bertin}\ \emph {et~al.}(2006)\citenamefont {Bertin}, \citenamefont {Droz},\ and\ \citenamefont {Gr{\'e}goire}}]{bertin2006boltzmann}%
  \BibitemOpen
  \bibfield  {author} {\bibinfo {author} {\bibfnamefont {E.}~\bibnamefont {Bertin}}, \bibinfo {author} {\bibfnamefont {M.}~\bibnamefont {Droz}},\ and\ \bibinfo {author} {\bibfnamefont {G.}~\bibnamefont {Gr{\'e}goire}},\ }\bibfield  {title} {\bibinfo {title} {Boltzmann and hydrodynamic description for self-propelled particles},\ }\href {https://doi.org/https://doi.org/10.1103/PhysRevE.74.022101} {\bibfield  {journal} {\bibinfo  {journal} {Physical Review E—Statistical, Nonlinear, and Soft Matter Physics}\ }\textbf {\bibinfo {volume} {74}},\ \bibinfo {pages} {022101} (\bibinfo {year} {2006})}\BibitemShut {NoStop}%
\bibitem [{\citenamefont {Duan}\ \emph {et~al.}(2024)\citenamefont {Duan}, \citenamefont {Agudo-Canalejo}, \citenamefont {Golestanian},\ and\ \citenamefont {Mahault}}]{duan2024phase}%
  \BibitemOpen
  \bibfield  {author} {\bibinfo {author} {\bibfnamefont {Y.}~\bibnamefont {Duan}}, \bibinfo {author} {\bibfnamefont {J.}~\bibnamefont {Agudo-Canalejo}}, \bibinfo {author} {\bibfnamefont {R.}~\bibnamefont {Golestanian}},\ and\ \bibinfo {author} {\bibfnamefont {B.}~\bibnamefont {Mahault}},\ }\bibfield  {title} {\bibinfo {title} {Phase coexistence in nonreciprocal quorum-sensing active matter},\ }\href {https://doi.org/10.48550/arXiv.2411.05465} {\bibfield  {journal} {\bibinfo  {journal} {arXiv preprint arXiv:2411.05465}\ } (\bibinfo {year} {2024})}\BibitemShut {NoStop}%
\bibitem [{\citenamefont {Seyed-Allaei}\ \emph {et~al.}(2016)\citenamefont {Seyed-Allaei}, \citenamefont {Schimansky-Geier},\ and\ \citenamefont {Ejtehadi}}]{seyed2016gaussian}%
  \BibitemOpen
  \bibfield  {author} {\bibinfo {author} {\bibfnamefont {H.}~\bibnamefont {Seyed-Allaei}}, \bibinfo {author} {\bibfnamefont {L.}~\bibnamefont {Schimansky-Geier}},\ and\ \bibinfo {author} {\bibfnamefont {M.~R.}\ \bibnamefont {Ejtehadi}},\ }\bibfield  {title} {\bibinfo {title} {Gaussian theory for spatially distributed self-propelled particles},\ }\href {https://doi.org/https://doi.org/10.1103/PhysRevE.94.062603} {\bibfield  {journal} {\bibinfo  {journal} {Physical Review E}\ }\textbf {\bibinfo {volume} {94}},\ \bibinfo {pages} {062603} (\bibinfo {year} {2016})}\BibitemShut {NoStop}%
\bibitem [{\citenamefont {Joanny}(2015)}]{grosberg2015nonequilibrium}%
  \BibitemOpen
  \bibfield  {author} {\bibinfo {author} {\bibfnamefont {J.-F.}\ \bibnamefont {Joanny}},\ }\bibfield  {title} {\bibinfo {title} {Nonequilibrium statistical mechanics of mixtures of particles in contact with different thermostats},\ }\href {https://doi.org/https://doi.org/10.1103/PhysRevE.92.032118} {\bibfield  {journal} {\bibinfo  {journal} {Physical Review E}\ }\textbf {\bibinfo {volume} {92}},\ \bibinfo {pages} {032118} (\bibinfo {year} {2015})}\BibitemShut {NoStop}%
\bibitem [{\citenamefont {Weber}\ \emph {et~al.}(2016)\citenamefont {Weber}, \citenamefont {Weber},\ and\ \citenamefont {Frey}}]{weber2016binary}%
  \BibitemOpen
  \bibfield  {author} {\bibinfo {author} {\bibfnamefont {S.~N.}\ \bibnamefont {Weber}}, \bibinfo {author} {\bibfnamefont {C.~A.}\ \bibnamefont {Weber}},\ and\ \bibinfo {author} {\bibfnamefont {E.}~\bibnamefont {Frey}},\ }\bibfield  {title} {\bibinfo {title} {Binary mixtures of particles with different diffusivities demix},\ }\href {https://doi.org/https://doi.org/10.1103/PhysRevLett.116.058301} {\bibfield  {journal} {\bibinfo  {journal} {Physical review letters}\ }\textbf {\bibinfo {volume} {116}},\ \bibinfo {pages} {058301} (\bibinfo {year} {2016})}\BibitemShut {NoStop}%
\bibitem [{\citenamefont {Damman}\ \emph {et~al.}(2024)\citenamefont {Damman}, \citenamefont {D{\'e}mery}, \citenamefont {Palumbo},\ and\ \citenamefont {Thomas}}]{damman2024algebraic}%
  \BibitemOpen
  \bibfield  {author} {\bibinfo {author} {\bibfnamefont {P.}~\bibnamefont {Damman}}, \bibinfo {author} {\bibfnamefont {V.}~\bibnamefont {D{\'e}mery}}, \bibinfo {author} {\bibfnamefont {G.}~\bibnamefont {Palumbo}},\ and\ \bibinfo {author} {\bibfnamefont {Q.}~\bibnamefont {Thomas}},\ }\bibfield  {title} {\bibinfo {title} {Algebraic depletion interactions in two-temperature mixtures},\ }\href {https://doi.org/10.48550/arXiv.2406.11616} {\bibfield  {journal} {\bibinfo  {journal} {arXiv preprint arXiv:2406.11616}\ } (\bibinfo {year} {2024})}\BibitemShut {NoStop}%
\bibitem [{\citenamefont {Dorigo}\ \emph {et~al.}(2013)\citenamefont {Dorigo}, \citenamefont {Floreano}, \citenamefont {Gambardella}, \citenamefont {Mondada}, \citenamefont {Nolfi}, \citenamefont {Baaboura}, \citenamefont {Birattari}, \citenamefont {Bonani}, \citenamefont {Brambilla}, \citenamefont {Brutschy} \emph {et~al.}}]{dorigo2013swarmanoid}%
  \BibitemOpen
  \bibfield  {author} {\bibinfo {author} {\bibfnamefont {M.}~\bibnamefont {Dorigo}}, \bibinfo {author} {\bibfnamefont {D.}~\bibnamefont {Floreano}}, \bibinfo {author} {\bibfnamefont {L.~M.}\ \bibnamefont {Gambardella}}, \bibinfo {author} {\bibfnamefont {F.}~\bibnamefont {Mondada}}, \bibinfo {author} {\bibfnamefont {S.}~\bibnamefont {Nolfi}}, \bibinfo {author} {\bibfnamefont {T.}~\bibnamefont {Baaboura}}, \bibinfo {author} {\bibfnamefont {M.}~\bibnamefont {Birattari}}, \bibinfo {author} {\bibfnamefont {M.}~\bibnamefont {Bonani}}, \bibinfo {author} {\bibfnamefont {M.}~\bibnamefont {Brambilla}}, \bibinfo {author} {\bibfnamefont {A.}~\bibnamefont {Brutschy}}, \emph {et~al.},\ }\bibfield  {title} {\bibinfo {title} {Swarmanoid: a novel concept for the study of heterogeneous robotic swarms},\ }\href {https://doi.org/10.1109/MRA.2013.2252996} {\bibfield  {journal} {\bibinfo  {journal} {IEEE Robotics \& Automation Magazine}\ }\textbf {\bibinfo {volume} {20}},\ \bibinfo {pages} {60} (\bibinfo {year}
  {2013})}\BibitemShut {NoStop}%
\bibitem [{\citenamefont {Yasuda}\ \emph {et~al.}(2014)\citenamefont {Yasuda}, \citenamefont {Adachi},\ and\ \citenamefont {Ohkura}}]{yasuda2014self}%
  \BibitemOpen
  \bibfield  {author} {\bibinfo {author} {\bibfnamefont {T.}~\bibnamefont {Yasuda}}, \bibinfo {author} {\bibfnamefont {A.}~\bibnamefont {Adachi}},\ and\ \bibinfo {author} {\bibfnamefont {K.}~\bibnamefont {Ohkura}},\ }\bibfield  {title} {\bibinfo {title} {Self-organized flocking of a mobile robot swarm by topological distance-based interactions},\ }in\ \href {https://doi.org/10.1109/SII.2014.7028020} {\emph {\bibinfo {booktitle} {2014 IEEE/SICE International Symposium on System Integration}}}\ (\bibinfo {organization} {IEEE},\ \bibinfo {year} {2014})\ pp.\ \bibinfo {pages} {106--111}\BibitemShut {NoStop}%
\bibitem [{\citenamefont {Marchetti}\ \emph {et~al.}(2013)\citenamefont {Marchetti}, \citenamefont {Joanny}, \citenamefont {Ramaswamy}, \citenamefont {Liverpool}, \citenamefont {Prost}, \citenamefont {Rao},\ and\ \citenamefont {Simha}}]{marchetti2013hydrodynamics}%
  \BibitemOpen
  \bibfield  {author} {\bibinfo {author} {\bibfnamefont {M.~C.}\ \bibnamefont {Marchetti}}, \bibinfo {author} {\bibfnamefont {J.~F.}\ \bibnamefont {Joanny}}, \bibinfo {author} {\bibfnamefont {S.}~\bibnamefont {Ramaswamy}}, \bibinfo {author} {\bibfnamefont {T.~B.}\ \bibnamefont {Liverpool}}, \bibinfo {author} {\bibfnamefont {J.}~\bibnamefont {Prost}}, \bibinfo {author} {\bibfnamefont {M.}~\bibnamefont {Rao}},\ and\ \bibinfo {author} {\bibfnamefont {R.~A.}\ \bibnamefont {Simha}},\ }\bibfield  {title} {\bibinfo {title} {Hydrodynamics of soft active matter},\ }\href {https://doi.org/10.1103/RevModPhys.85.1143} {\bibfield  {journal} {\bibinfo  {journal} {Rev. Mod. Phys.}\ }\textbf {\bibinfo {volume} {85}},\ \bibinfo {pages} {1143} (\bibinfo {year} {2013})}\BibitemShut {NoStop}%
\end{thebibliography}%
	
\end{document}